\documentclass[global,final]{svjour}
\usepackage[logonly]{trace}
\usepackage{color}
\usepackage{soul}
\long\def\comment#1{}

\usepackage[
  pdftitle={How to Compare the Scientific Contributions between Research Groups},
  pdfauthor={Hyoungshick Kim, Ji Won Yoon}, 
  pdfkeywords={Publication Analysis, Publication Trend, Comparative Analysis, Network Analysis, Security Research Trend},
  pdfborder={0 0 0},
  ]{hyperref}
  
\usepackage[bottom]{footmisc}
\usepackage{amsmath,epsfig}
\usepackage{natbib}

\usepackage{graphics}
  
\title{How to Compare the Scientific Contributions between Research Groups}

\author{Hyoungshick Kim\inst{1}\and Ji Won Yoon\inst{2}}

\institute{Computer Laboratory, University of Cambridge, UK, \email{hk331@cl.cam.ac.uk} \and Statistics department, Trinity College Dublin, Ireland, \email{yoonj@tcd.ie}}

\date{22 October 2009}

\titlerunning{How to Compare the Scientific Contributions between Research Groups}

\journalname{Scientometrics}

\begin{document}

\maketitle

\begin{abstract}
We present a method to analyse the scientific contributions between research groups. Given multiple research groups, we construct their journal/proceeding graphs and then compute the similarity/gap between them using network analysis. This analysis can be used for measuring similarity/gap of the topics/qualities between research groups' scientific contributions. We demonstrate the practicality of our method by comparing the scientific contributions by Korean researchers with those by the global researchers for information security in 2006 -- 2008. The empirical analysis shows that the current security research in South Korea has been isolated from the global research trend.
\end{abstract}

\section{Introduction}
\label{sec:intro}
For many areas, the evaluation of scientific contributions is a significant issue in the allocation of research funding and the assessment of the quality of research conducted by universities, institutes or countries.

Peer review, where evaluation process is based on judgements formulated by independent experts, is commonly accepted as an ideal solution for this purpose since scientific contribution can be effectively evaluated by experts who are knowledgeable in the subject area being reviewed. Rankings and supporting qualitative evaluations by the experts can provide comparative information between research groups. However, despite its desirable effectiveness, peer review has a troublesome and challenging task in practice; this is how to assign unbiased and transparent experts. Surely, it is not trivial to recruit peer review committees who are composed of specialists related to a particular subject on time and within a limited budget~\cite{Seglen97:Education}. Moreover, we note that peer review is relatively slow and inefficient to reach a final decision. 

Alternatively, it has been tempting to use bibliometrics as simple and practical tools to assess scientific contribution. Bibliometric indicators such as the number of publications, journal impact factors, number of citations, and citation index can be readily available and also provide some meaningful information on the level of research productivity and scientific impact. Not surprisingly, it is really important to use a bibliometric database which is suitable for a purpose since these indicators can be greatly changed depending on the bibliometric database being used. 

The ISI bibliographic database, which includes the Arts and Humanities Citation Index (A\& HCI), Science Citation Index (SCI) and Social Sciences Citation Index (SSCI), has been used for decades as de facto standard databases for conducting publication and citation analyses~\cite{Meho06:ResearchEvaluation}. However, it is not desirable to view this as universal database regardless of the purpose. First of all, the coverage of the database is not complete according to subjects. Different research fields are covered unequally and only a few of conference proceedings and books, which are also important scientific literatures, are included in the database. Unlike the other fields such as natural sciences and life sciences, prestigious conferences hosted by professional computer science societies such as ACM/IEEE are preferred to journals as a place to present original and important results~\cite{Fortnow09:ResearchEvaluation, Meyer09:ReserachEvaluation}. Moreover, some national journals, which are important in the social sciences and humanities, may not be considered since the databases have an English language bias~\cite{Seglen98:ResearchEvaluation}. Lastly, although the database attempts to include the most important scientific literatures for a specific subject, it is difficult to estimate the only scientific contributions relevant to the specific subject since other unrelated literatures are also included in the database. For example, suppose that we want to evaluate a research group's the scientific contributions to Russian history. The ISI bibliographic database is not proper for this purpose since some relevant (Russian) literatures may not be included in the database, whereas unnecessary literatures can be included. Our study is motivated by this limitation of the dependency on the bibliographic database.

Our goal is to design a research evaluation method, which can compare the scientific contributions of research groups directly, without a specific bibliographic database. We propose how to compare the scientific contributions of research groups, inspired by recent advances in complex network analysis. This analysis can be a good alternative to the peer review or the conventional bibliometric indicators since we can compare the scientific contributions of a given research group with well-known experts' scientific contributions. In this paper, we make the following two contributions.
\begin{itemize}
\item We propose an analysis method to measure the similarity/gap between research groups by comparing their publication patterns in Section \ref{sec:The proposed method}. For comparison of publication patterns, we construct the relationship graphs on their publications and then analyse the relevance between the constructed graphs. We suggest the metrics to measure the similarity/gap between the research groups' publication outputs. This method is useful to see how much close to the research mainstream in a specific field.
\item As a practical application, we compare the publication outputs of South Korea with those of the global researchers for information security during the period 2006 -- 2008 in Section \ref{sec:An example}. Our main results are shown in Table \ref{tab: overlapping nodes of the projected graphs} and \ref{tab: distance between the projected graphs} in Section \ref{sec:Comparison of two groups}. The experimental result shows that as suspected, Korean security researchers have been somewhat isolated from the mainstream.
\end{itemize}

Although the proposed measurement does not mean the research quality of the scientific contributions from a research group, this analysis can measure how much the publication outputs of a research groups is similar to those of another research group. Consequently, it can be applied to a useful supplement for research evaluation or trend analysis. 

\section{The proposed method}
\label{sec:The proposed method}

Our goal is to analyse the similarity/gap between the scientific contributions of the multiple research groups. Firstly, we construct each research group's journal/proceeding graph using their publication outputs and then analyse the similarity/gap between them by comparing the constructed graphs.

If we compare the sets of researchers with different cardinalities, appropriate normalization is required. For simplicity, we assume that all the sets being compared have the same cardinality. 

\subsection{Construction of journal/proceeding graphs}
\label{sec:Construction of journal graphs}

Given a set of researchers ${\bf R}$, we construct the journal/proceeding graph ${\bf G}_{\bf R}^{\bf J}$ by taking the following steps:

\begin{enumerate}
\item For each researcher $a\in{\bf R}$, collect the $a$'s publication outputs within a time window (e.g. within 2008).
\item Generate the bipartite graph ${\bf G}_{\bf R}$ with these collected publication data, whose nodes are divided into a set of authors ${\bf A}$ and a set of journals/proceedings ${\bf J}$ and an edge $(a, j)$ means that the author $a$ published a paper in the journal (or proceeding) $j$ for $a\in{A}$ and $j\in{\bf J}$.
\item Construct the ${\bf J}$-projected graph ${\bf G}_{\bf R}^{\bf J}$ compressed by ${\bf J}$-projection, which is a well-known technique so-called one-mode projecting to show the relations among a particular set of nodes only~\cite{Guillaume05:Bipartite, Zhou07:Bipartite}. The ${\bf J}$-projection means a network containing only nodes in ${\bf J}$, where two nodes are connected when they have at least one common author. The weight of each edge is computed as 1/(the number of the shared authors).
\end{enumerate}

The constructed journal/proceeding graph may give the information about not only a set of topologically popular journals/proceedings for a research group but also the relative importance of them by computing their centrality metric values such as \emph{degree}, \emph{closeness} and \emph{betweenness}. We describe the definition and meaning of the metrics in Appendix \ref{appendix: Centrality metrics}. We denote ``$m$-central nodes'' the set of nodes of which $m$ metric values are greater than the average value of the graph. Consequently, we can identify the relatively important journals (or proceedings) in a journal/proceeding graph by observing the $m$-central nodes in the graph.

\begin{figure}[h!]
\centering
\begin{tabular}{c c}
\includegraphics[scale=0.25]{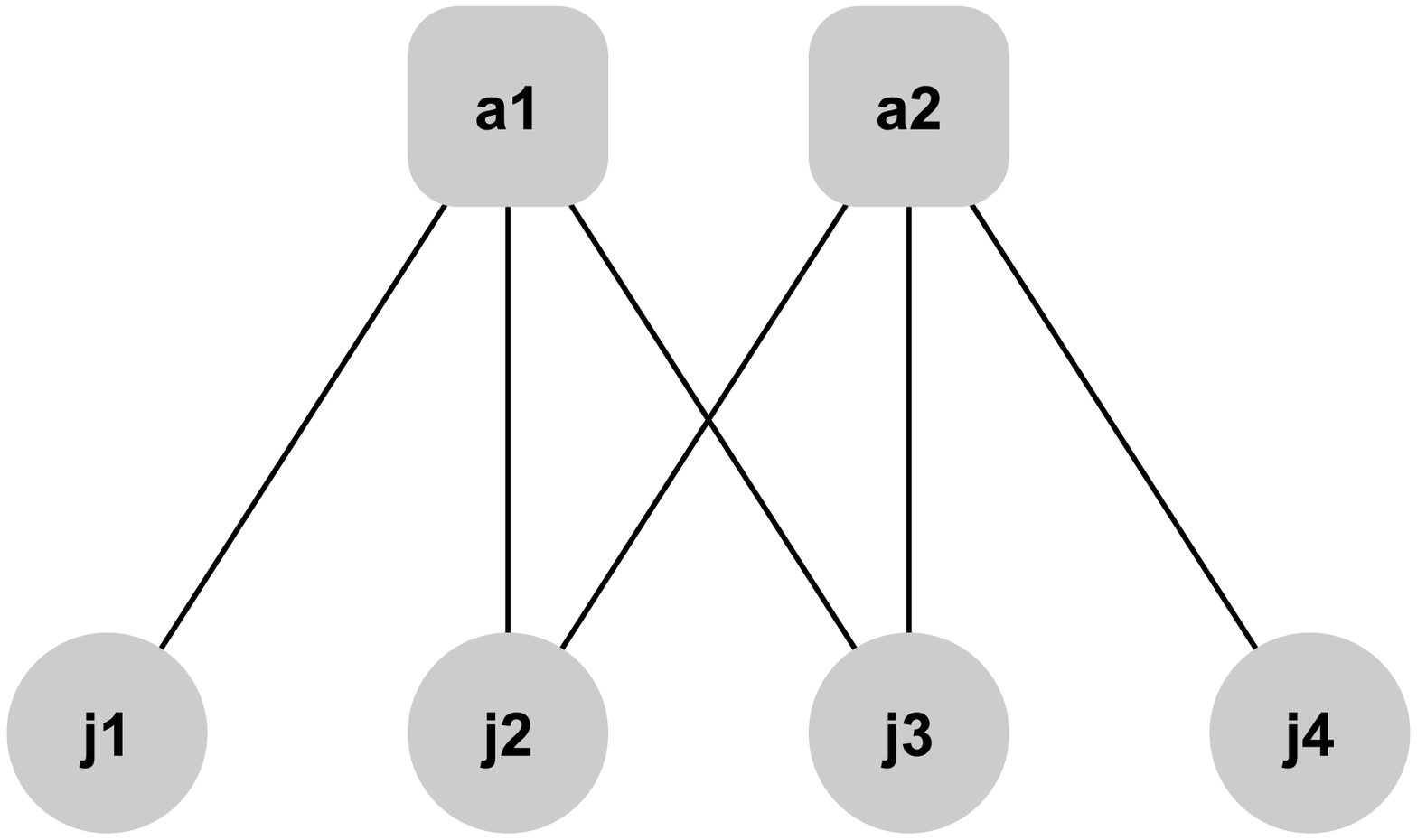} &
\includegraphics[scale=0.25]{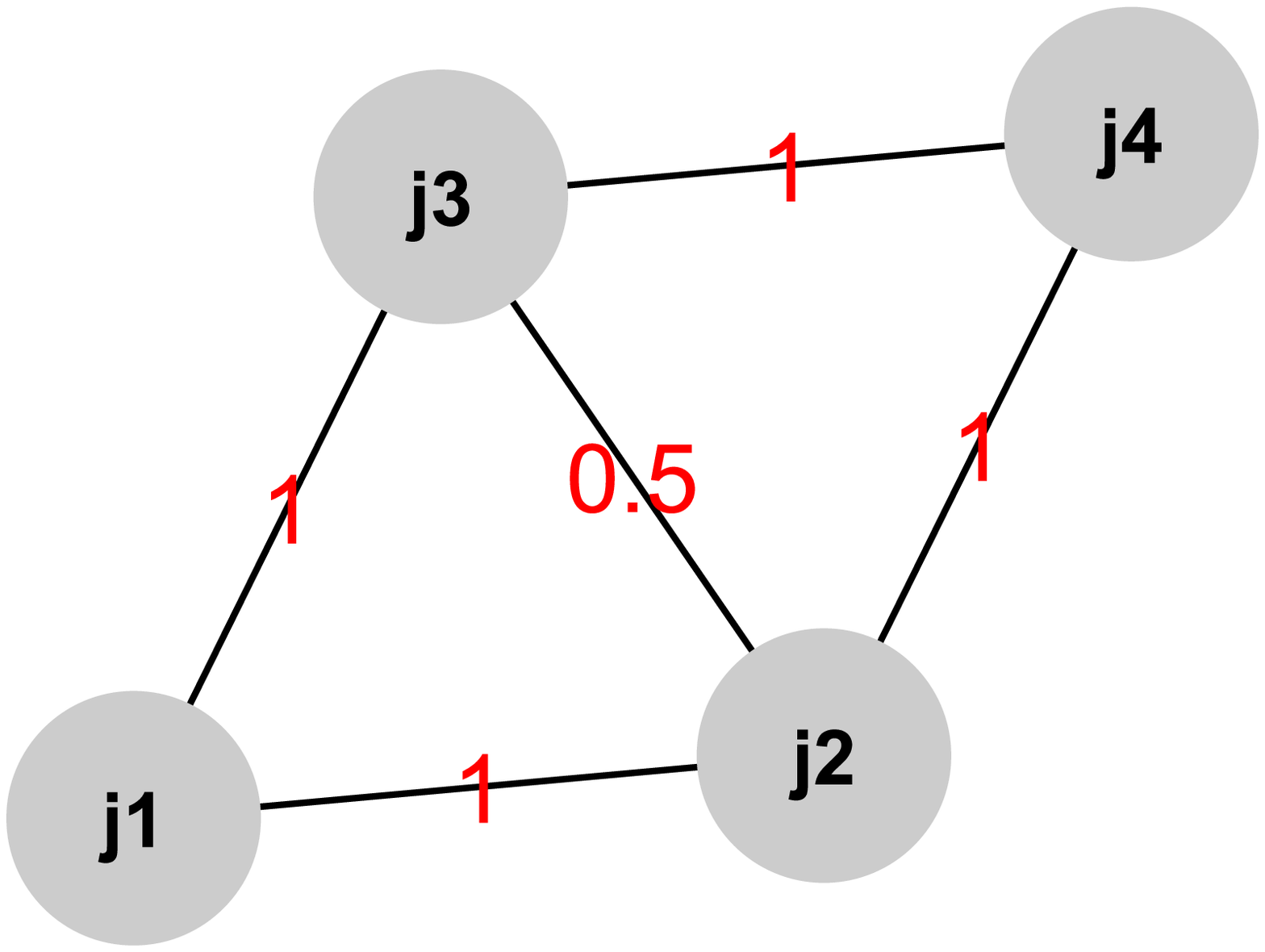}\cr
(a) A bipartite graph&
(b) A ${\bf J}$-projected graph
\end{tabular}
\caption{An example of journal/proceeding graph construction}
\label{fig: an example of journal graph}
\end{figure}

For an example, suppose that we have two researchers: ${\bf R}$ = $\{a_1, a_2\}$. When the researchers $a_1$ and $a_2$ published their papers in the journals ${j_1, j_2, j_3}$ and ${j_2, j_3, j_4}$, respectively, then we have a bipartite graph as shown in Figure \ref{fig: an example of journal graph} (a). From the bipartite graph, we can construct the ${\bf J}$-projected graph ${\bf G}_{\bf R}^{\bf J}$ as in Figure \ref{fig: an example of journal graph} (b). 
In our projecting method, the weight of each edge is assigned to be inversely proportional to \emph{the number of the shared authors} between two journals (or proceedings) to represent how to close them so that the weight of the edge $(j_2, j_3)$ is 0.5 since $a_1$ and $a_2$ are commonly published their papers in both journals $j_2$ and $j_3$. From this graph, we can identify $\{j_2, j_3\}$ as $degree$-central nodes of which degree values are greater than 2.

However, it is still rather difficult to explain the similarity/gap between the graphs although the nodes' centrality values show their relative importance for a research group. Therefore we need to define the metrics to measure the similarity/gap between the graphs quantitatively.

\subsection{Comparison of the journal/proceeding graphs}
\label{sec:Comparison of the journal graphs}

We analyse the similarity/gap between the journal/proceeding graphs constructed in Section \ref{sec:Construction of journal graphs}. For this purpose, we suggest the functions to measure the similarity/gap between networks explicitly. We classified these functions into two types: ``the fraction of overlapping nodes/interactions'' for similarity and ``the distance between the graphs'' for gap.

Given $k$ ${\bf J}$-projected graphs ${\bf G}_{1}^{\bf J} = ({V}_{1}, {E}_{1}), \cdots, {\bf G}_{k}^{\bf J} = ({V}_{k}, {E}_{k})$, the symbols $V_U$ and $V_A$ represent the superset of nodes in the graphs ($\bigcup_{i=1}^{k} V_i$) and the set of common nodes between the graph ($\bigcap_{i=1}^{k} V_i$), respectively. 

\subsubsection{Fraction of overlapping nodes/interactions}
\label{sec:Fraction of overlapping nodes}

To measure the similarity between the journal/proceeding graphs, we can simply count the number of common nodes or edges. The first metric is to compute the ratio of common nodes between the ${\bf J}$-projected graphs. We define this as follows:
\begin{definition}
\textbf{Ratio of common nodes}-- Given $k$ ${\bf J}$-projected graphs ${\bf G}_{1}^{\bf J} = ({V}_{1}, {E}_{1}), \cdots, {\bf G}_{k}^{\bf J} = ({V}_{k}, {E}_{k})$, we then define ``the ratio of common nodes'' as 
\begin{equation}
\text{\bf R}_\text{node}({\bf G}_{1}^{\bf J}, \cdots, {\bf G}_{k}^{\bf J}) = \frac{|V_A|}{|V_U|} \nonumber
\end{equation}
\label{def:Ratio of common nodes}
\end{definition}

The second metric is to compute the ratio of common interactions between the ${\bf J}$-projected graphs. Unlike Definition \ref{def:Ratio of common nodes}, however, we cannot simply achieve this goal since such a notion does not account for interactions of non-existing edges and is therefore not a comprehensive view of the interactions between the graphs.  

To represent non-existing edges, we define the function $e_i:V_U \times V_U \rightarrow \{0, 1\}$ by 
\begin{equation}
e_i(u, v) = \begin{cases} 1, & \mbox{if } (u, v)\in E_i \mbox{ ,}\\ 0, & \mbox{Otherwise .} \end{cases} \nonumber
\end{equation}
Likewise, we define the function $\bar{e}_{i}$ as the negation of $e_i$. Finally, we define the ratio of common interactions between the graphs with the functions $e_i$ and $\bar{e}_{i}$ as follows:
\begin{definition}
\textbf{Ratio of common interactions}-- Given $k$ ${\bf J}$-projected graphs ${\bf G}_{1}^{\bf J} = ({V}_{1}, {E}_{1}), \cdots, {\bf G}_{k}^{\bf J} = ({V}_{k}, {E}_{k})$, we denote $n$ the cardinality of $V_U$, and then define ``the ratio of common interactions'' as 
\[
\text{\bf R}_{\text{interaction}}({\bf G}_{1}^{\bf J}, \cdots, {\bf G}_{k}^{\bf J}) = \frac{2}{n(n-1)} \sum_{\substack{u, v \in V_U\\u\neq v}}\left( \prod_{i=1}^{k} {e}_{i}(u, v) + \prod_{i=1}^{k} {\bar{e}_{i}(u, v)}\right)
\]
\label{def:Ratio of common interactions}
\end{definition}

These metrics explain how much network structures (e.g. interaction patterns) are similar.

\subsubsection{Distance between the graphs}
\label{sec:Distance between the graphs}

In general, the computation of the common parts between graphs may not be suitable for comparison of journal/proceeding graphs since they have a few common nodes and edges. As more sophisticated measures, we can consider the distance between the nodes in the ${\bf J}$-projected graphs. The first metric is to compute the distance between the common nodes and the other nodes of the ${\bf J}$-projected graphs. We define this as follows:

\begin{definition}
\textbf{Closeness of common nodes}-- Given $k$ ${\bf J}$-projected graphs ${\bf G}_{1}^{\bf J} = ({V}_{1}, {E}_{1}), \cdots, {\bf G}_{k}^{\bf J} = ({V}_{k}, {E}_{k})$, we define ``the closeness of common nodes'' as follows:
\begin{equation}
\text{C}_{\text{common}}({\bf G}_{1}^{\bf J}, \cdots, {\bf G}_{k}^{\bf J}) = \begin{cases} \frac{1}{|V_U|}\cdot \displaystyle\sum_{u \in V_U} \displaystyle\min_{v \in V_A} distance(u, v) , & \mbox{if } V_A \neq \emptyset \mbox{ ,}\\ \infty , & \mbox{Otherwise .} \end{cases} \nonumber
\end{equation}
\end{definition}

This metric measures how close all other journals/proceedings in the network are located from their common journal/proceedings. We can explain how much closer a node in each graph to the common nodes between the graphs on the average using this value. This value will be exactly 0 if and only if $V_A$ is the same as $V_U$. 

For some applications, it is also important to observe the diversity of journals/proceedings between researchers. Basically, this property is closely related to the \emph{network diameter}\footnote{Network diameter is the maximum distance between nodes in the network~\cite{Hage95:SocialNetwork}} of a ${\bf J}$-projected graph. Therefore we need to measure how many the \emph{network diameter} of the union graph ${\bf G}_{U}^{\bf J} = ({V}_{U}, {E}_{U})$ is increased after combining all the ${\bf J}$-projected graphs where ${E}_{U} = \bigcup_{i=1}^{k} E_i$. We compute the average increasing size of the union graph ${\bf G}_{U}^{\bf J}$ as follows:

\begin{definition}
\textbf{Average increasing diameter}-- Given $k$ ${\bf J}$-projected graphs ${\bf G}_{1}^{\bf J} = ({V}_{1}, {E}_{1}), \cdots, {\bf G}_{k}^{\bf J} = ({V}_{k}, {E}_{k})$, we define ``the average increasing diameter'' as follows:
\begin{equation}
\Delta\text{D}({\bf G}_{1}^{\bf J}, \cdots, {\bf G}_{k}^{\bf J}) = \frac{1}{k}\cdot \displaystyle\sum_{1=1}^{k} (diameter(G_U)-diameter(G_i)) \nonumber
\end{equation}
\end{definition}

\section{An example}
\label{sec:An example}

We demonstrate the practicality of our method by comparing the scientific contributions by Korean researchers with those by the global researchers for information security in 2006 -- 2008. 

Our goal is to show how much closer the scientific contributions by Korean researchers to the research mainstream by comparing their publication outputs with the well known global researchers' results. As an example, we analyse security research in South Korea from 2006 to 2008. We use a sample set since it is practically infeasible to collect all publications related to security. To obtain a reasonable sample, we perform the following two steps:
\begin{enumerate}
\item Select top conferences related to security field and held in South Korea. 
\item Randomly select $n$ Korean researchers from the program committee members of the selected conferences in South Korea.
\end{enumerate}
In selecting conferences, some prior knowledge is required. We select two conferences, ``International Workshop on Information Security Applications'' (WISA) \cite{WISA:WebSite} and ``International Conference on Information Security and Cryptology'' (ICISC) \cite{ICISC:WebSite}, on the basis of their large scale and long history compared to other conferences. Also, we define the sample size as 20 ($n=20$). We assume that 20 active researchers are enough to show the characteristics or trend. Let $T$ be a set of randomly selected researchers from the program committee members of these conferences.

In the similar manner, we obtain a reasonable sample set of global researchers by using the top international conferences for security, ``IEEE Symposium on Security and Privacy'', ``ACM Conference on Computer and Communications Security'' and ``Usenix Security Symposium''. These conferences are selected under the conference ranking of well-known web sites \cite{SCR:Ranking, CSCR:Ranking, CSCRS:Ranking}. Let $P$ be a set of randomly selected researchers from the program committee members of these conferences.

We collect $T$'s and $P$'s publication results from 2006 to 2008, respectively. For simplicity, we only consider the bibliographic information indexed by the Digital Bibliography \& Library Project (DBLP) \cite{Ley02:Bibliography} under the assumption that this database provides the most bibliographic information on major computer science journals and proceedings.

\begin{table}
\begin{center}
\begin{tabular}{||r||c|c|c|c||}
\cline{2-4}\cline{2-4}
   \multicolumn{1}{c||}{} & $|A|$ & $|J|$ & $\sharp$ journals/proceedings \\
  \hline\hline
  2006 (Korean) & 14 & {\bf 40} & 77 \\
  2007 (Korean) & 16 & {\bf 43} & 61 \\
  2008 (Korean) & 14 & {\bf 37} & 51 \\\hline
  2006 (Global) & 14 & {\bf 58} & 73 \\
  2007 (Global) & 16 & {\bf 46} & 59 \\  
  2008 (Global) & 17 & {\bf 55} & 68 \\
  \hline\hline
\end{tabular}
\end{center}
\caption{Summary of publication data: $|A|$ and $|J|$ represent the number (cardinality) of the authors and their publications, respectively. We have $|A|<20$ since we plotted only authors who has at least one publication.}
\label{tab: summary of publication data}
\end{table}

With the collected publication data, we construct the bipartite graphs for each year and each research group from 2006 to 2008. From these bipartite graphs, we analyse the basic network properties which we summarize in Table \ref{tab: summary of publication data}. Since some researchers in $T$ and $P$ do not have any publications in the DBLP database during 2006 -- 2008, we can only draw between 14 and 17 authors who have at least one publication in the related year among 20 sampled researchers.

\subsection{Journal/proceeding graphs}
\label{sec:journal graphs}
\begin{figure*}[!h]
\centering
\begin{tabular}{c c}
\includegraphics[scale=0.28]{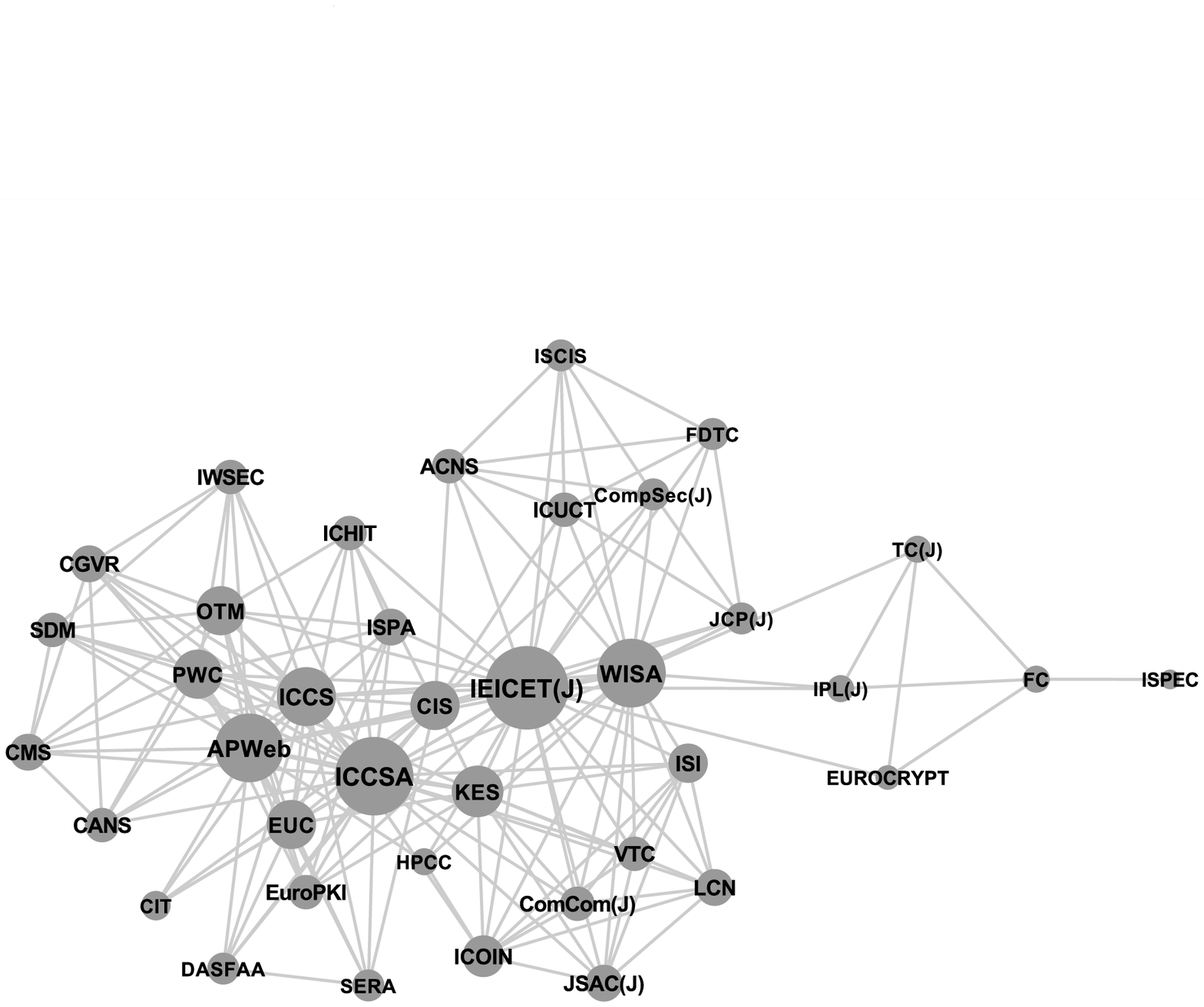}&
\includegraphics[scale=0.28]{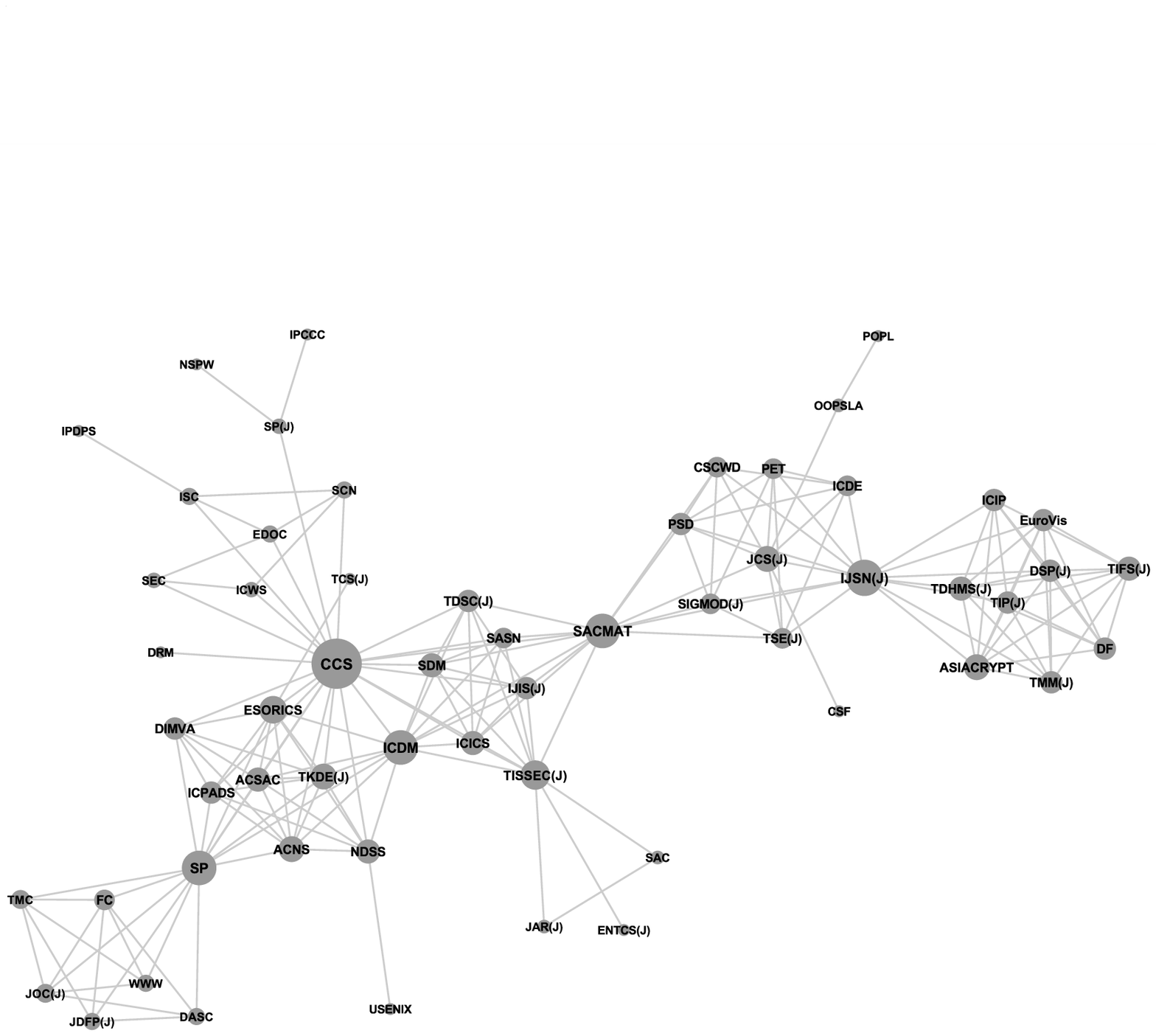}\cr
(a) Korean researchers (2006)&
(b) Global researchers (2006)\cr
\includegraphics[scale=0.28]{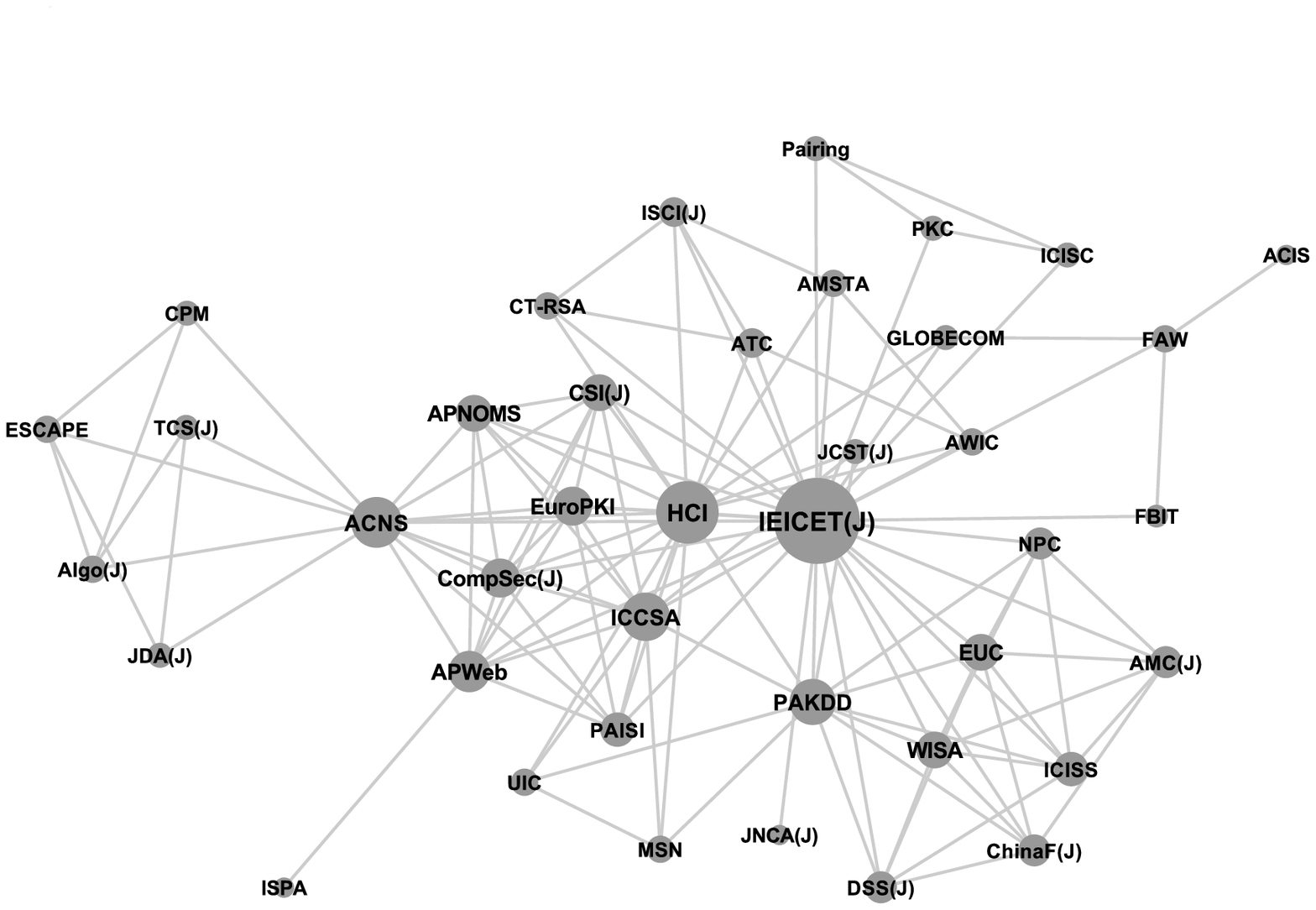}&
\includegraphics[scale=0.28]{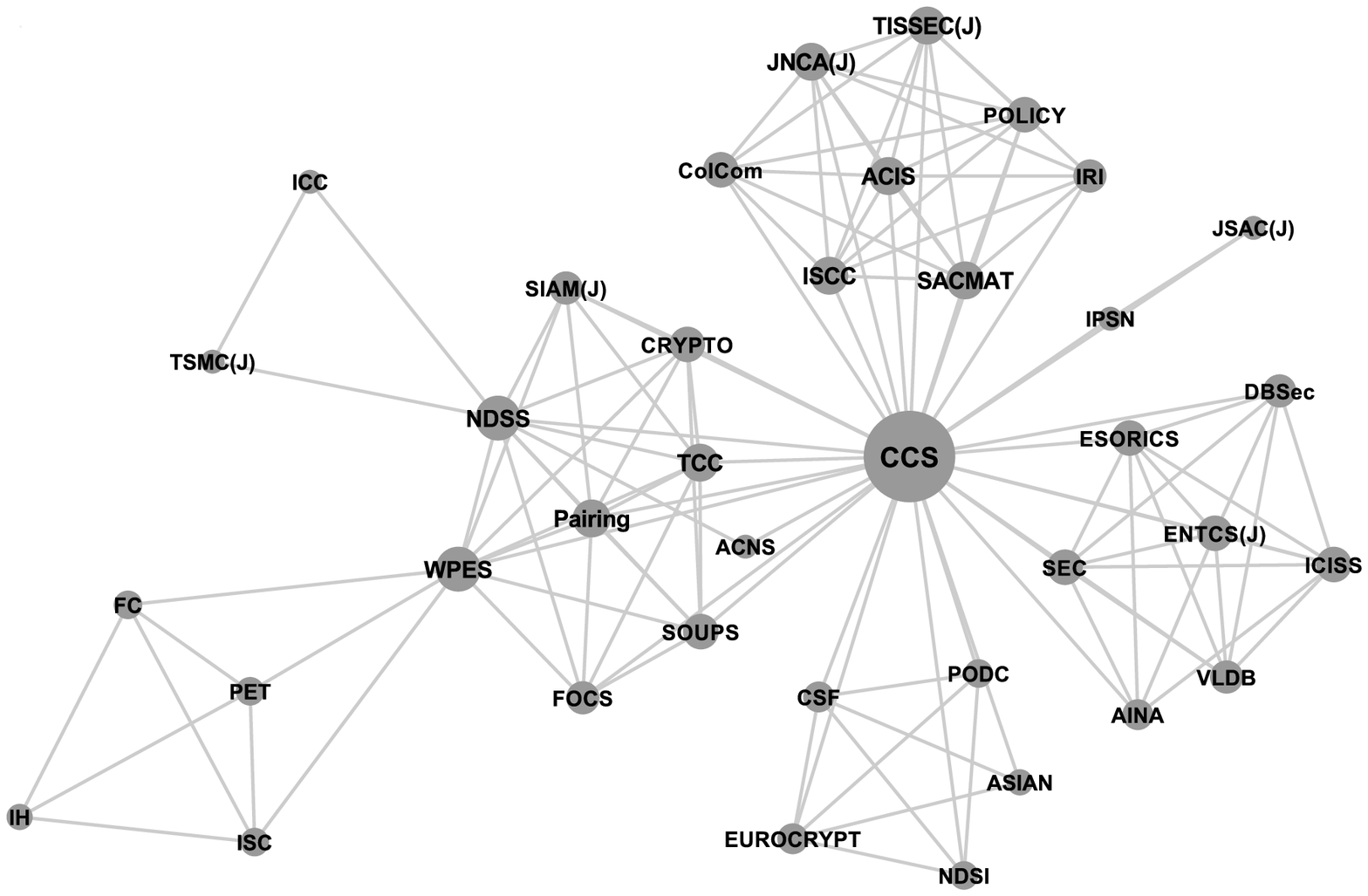}\cr
(c) Korean researchers (2007)&
(d) Global researchers (2007)\cr
\includegraphics[scale=0.28]{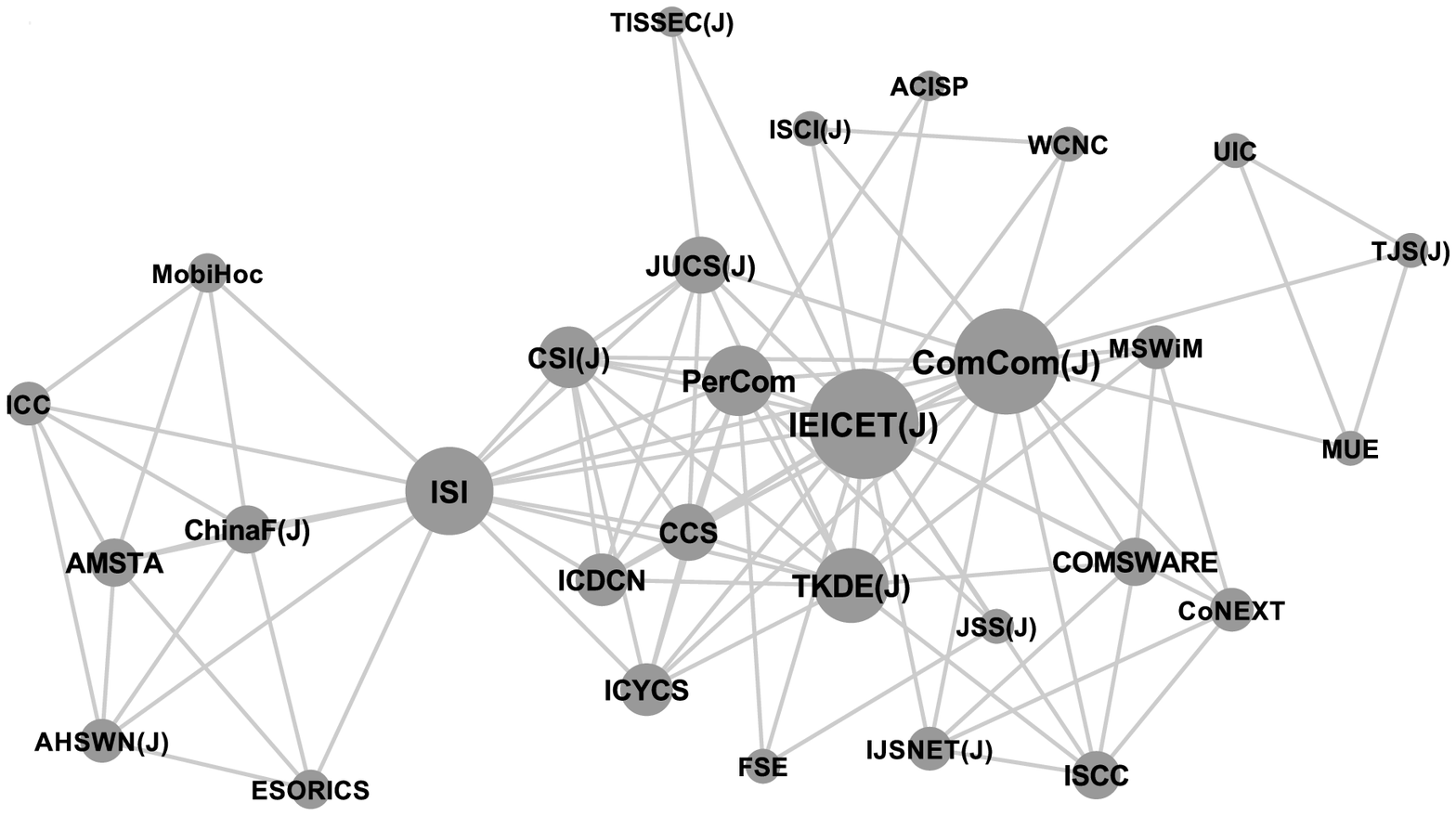}&
\includegraphics[scale=0.28]{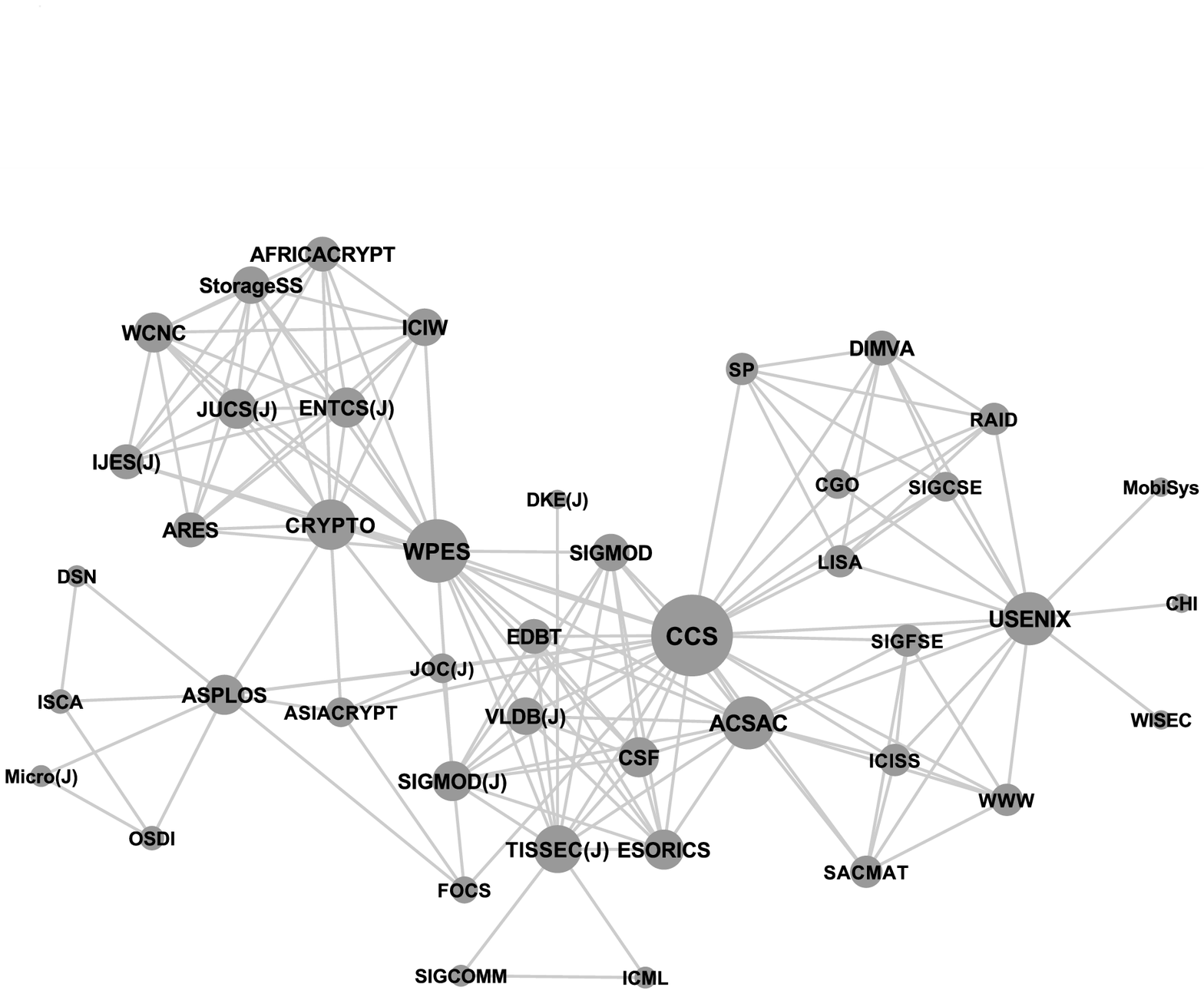}\cr
(e) Korean researchers (2008)&
(f) Global researchers (2008)\cr
\end{tabular}
\caption{${\bf J}$-projected graphs from 2006 to 2008}
\label{fig: projected graphs between two research groups}
\end{figure*}

By ${\bf J}$-projection in Section \ref{sec:Construction of journal graphs}, we construct ${\bf J}$-projected graphs\footnote{Without loss of generality, in the case of a disconnected graph, we only consider the largest connected component in the graph since it is commonly believed that the largest component is most meaningful in practice.} from the bipartite graphs. The resulting graphs are shown in Figure \ref{fig: projected graphs between two research groups}. In Figure \ref{fig: projected graphs between two research groups}, the size of each node is increased to be linearly proportional to the node's degree and the acronyms of journals/proceedings (We provide a supplementary material to introduce the full name of journals/proceedings\footnote{In the acronyms, (J) means a journal article.}) are used as nodes' identifiers.

We summarize several basic network properties of each ${\bf J}$-projected graph in Table \ref{tab: graphs structural properties}. We observe that the network size of Korean researchers' ${\bf J}$-projected graphs appear to be approximately decreasing from year to year.
\begin{table}[h!]
\begin{center}
\begin{tabular}{||r||c|c|c|c|c|c|c||}
\cline{2-5}\cline{2-5}
   \multicolumn{1}{c||}{} & $\sharp$ nodes & $\sharp$ edges & average & diameter \\
   \multicolumn{1}{c||}{} & & & distance & 		 \\
  \hline\hline
  2006 (Korean) & 39 & 180 & 2.000 & 5 \\
  2007 (Korean) & 40 & 128 & 2.103 & 4 \\
  2008 (Korean) & 30 & 99 & 1.993 & 3 \\\hline
  2006 (Global) & 58 & 181 & 3.019 & 6 \\
  2007 (Global) & 38 & 123 & 2.077 & 4 \\  
  2008 (Global) & 44 & 155 & 2.297 & 4 \\
  \hline\hline
\end{tabular}
\end{center}
\caption{${\bf J}$-projected graphs' properties}
\label{tab: graphs structural properties}
\end{table}

As we discussed in Section \ref{sec:Construction of journal graphs}, we can interpret the relative prominence of journals/proceedings embedded in the graphs by computing centrality metric values such as \emph{degree}, \emph{closeness} and \emph{betweenness}. We can identify the $m$-central nodes for each research group from Figures \ref{fig: High degree nodes in the projected graphs}, \ref{fig: High closeness nodes in the projected graphs} and \ref{fig: High betweenness nodes in the projected graphs} in Appendix \ref{appendix: Central journals/proceedings in the example}. While we can see that each research group's central journals/proceedings have not changed very much over time, there is almost no common central node between ``Korean researchers''' and ``Global researchers''' graphs. In particular, ``IEICE(J)'' and ``CCS'' is the key journal (or proceeding) for ``Korean research group'' and ``Global research group'', respectively. 

\subsection{Comparison of two research groups}
\label{sec:Comparison of two groups}

First of all, we measure the metrics of overlapping nodes/interactions. The results are shown in Table \ref{tab: overlapping nodes of the projected graphs}. 
\begin{table}[h!]
\begin{center}
\begin{tabular}{||r|c|c|c||}
\hline\hline
   & Common nodes & ${\bf R}_{node}$ & ${\bf R}_{interaction}$ \\
  \hline
  2006 & ACNS, FC, SDM & \bf{0.032} & 0.472\\
  2007 & ACIS, ACNS, ICISS, JNCA(J), Pairing & \bf{0.068} & 0.472\\
  2008 & CCS, ESORICS, JUCS(J), TISSEC(J), WCNC & \bf{0.072} & 0.466\\
  \hline\hline
\end{tabular}
\end{center}
\caption{Overlapping nodes/edges between the ${\bf J}$-projected graphs}
\label{tab: overlapping nodes of the projected graphs}
\end{table}
While all ``the ratios of the common nodes'' are under 10\%, all ``the ratios of the common interactions'' are higher than the 45\%. We note that ``the ratios of the common interactions'' is not a meaningful metric in this application since the graphs are too sparse and there is no common edge between two research groups' journal/proceeding graphs.

For measuring the gap between these graphs, we compute the metric values in Section \ref{sec:Distance between the graphs}. The results are shown in Table \ref{tab: distance between the projected graphs}.

\begin{table}[h!]
\begin{center}
\begin{tabular}{||r|c|c||}
\hline\hline
   & $\text{C}_{\textrm{common}}$ & $\Delta\text{D}$ \\
  \hline
  2006 & \bf{1.709} & 1.5\\
  2007 & \bf{1.288} & 2.0\\
  2008 & \bf{1.105} & 0.5\\
  \hline\hline
\end{tabular}
\end{center}
\caption{Distance between the ${\bf J}$-projected graphs \label{tab: distance between the projected graphs}}
\end{table}

From Table \ref{tab: distance between the projected graphs}, we can see that the distance between the common nodes and the other nodes of the ${\bf J}$-projected graphs is continuously decreased from year to year. We can also see that $\Delta D$ is nearly close to 0 since the network diameter of the union graph ${\bf G}_{U}^{\bf J}$ is decreased to 4 in 2008. This shows that the journals/proceedings which the Korean researchers submitted to do not deviate much from the research mainstream.

\subsection{Discussion}
\label{sec:Discussion}

We compared the journals/proceedings that Korean researchers have focused on with those that global researchers have focused on by projecting the bipartite graphs into projected graphs. In Table \ref{tab: overlapping nodes of the projected graphs} and \ref{tab: distance between the projected graphs}, we found that the Korean and global research groups share only a small fraction of journals/proceedings and their journal/proceeding graphs have somewhat different structures. That is, Korean researchers and global researchers are publishing their papers in different journals or conferences even though they are working in the same subject. Under the assumption that a global research group is close to the ideal research group, we claim that the Korean research group will have to exert itself more than it currently does to publish many papers in journals/proceedings with high centrality (e.g. CCS, WPES, USENIX and ACSAC) in the global researchers' graphs as shown in Figures \ref{fig: High degree nodes in the projected graphs}, \ref{fig: High closeness nodes in the projected graphs} and \ref{fig: High betweenness nodes in the projected graphs} in Appendix \ref{appendix: Central journals/proceedings in the example}. However, the metrics in Table \ref{tab: overlapping nodes of the projected graphs} and \ref{tab: distance between the projected graphs} also show that Korean security researchers' publication pattern in 2008 is somewhat close to the mainstream compared to that in 2006.

Our work is primarily intended to demonstrate how to compare publication patterns between the research groups. We have not considered research quality since the results of our metrics may not give enough evidence to compare quality between two groups (although we can guess). The proposed analysis of publication pattern can be, however, a useful supplement rather than a replacement for traditional research evaluation methods. 

In addition conference (or journal) selection is strongly related to geographical and political factors in the real world. In this paper, we do not consider these factors.

\section{Related work}

The use of statistical bibliometric indicators in research evaluation emerged in the 1960s and 1970s \cite{Leydesdorff05:ResearchEvaluation}, and is in wide use today due to the development of the relevant databases. These indicators provide useful output measures of activity and performance in scientific research and have become standard tools for research evaluation \cite{Almeida09:ResearchEvaluation}. However, some methodological problems of research evaluation at the micro level (e.g. the scientific contribution of a small research group) still remain unresolved \cite{Phelan99:ResearchEvaluation, Moed05:ResearchEvaluation}. Meyer et al. \cite{Meyer09:ReserachEvaluation} issued the problems of the bibliometric indicators for computer science in detail.

An alternative approach is to analyse researchers' social networks such as co-citation networks \cite{Chen99:ResearchEvaluation, Erten03:ResearchEvaluation, An04:ResearchEvaluation, Sidiropoulos05:ResearchEvaluation} and co-authorship networks \cite{Nascimento03:ResearchEvaluation, Hassan04:ResearchEvaluation, Horn04:ResearchEvaluation, Elmacioglu05:ResearchEvaluation}. Citation networks can be also used to evaluate the importance of journals/proceedings by computing centrality values of the nodes in a citation graph. Co-authorship networks are an important class of social networks and have been used extensively. Many co-authorship networks have been studied to investigate the patterns, motivation, and the structure of scientific collaboration \cite{Batagelj00:SocialNetwork, Egghe00:SocialNetwork, Newman01:ComplexNetwork, Newman01_2:ComplexNetwork, Farkas02:SocialNetwork, Newman04:ComplexNetwork, Kretschmer04:SocialNetwork, Liu05:SocialNetwork}. 
Morris \cite{Morris05:SocialNetwork} proposed a model to monitor the birth and development of a scientific speciality with a collection of journal papers. Lee \cite{Lee08:SocialNetwork} practically analysed the research trends in the information security field using ``co-word analysis''. Our work is to extend these to measure the similarity/gap between research groups by comparing their publication outputs.

\section{Conclusion}

We have presented a set of metrics to compare research groups' publication outputs and have shown how they can be applied effectively to measure the similarity/gap between them. For example, our proposed method can explain a research group's connectedness to the research mainstream, both statically and over time. We showed the similarity/gap between the publication patterns of Korean researchers and global in information security from 2006 to 2008. The experimental results show that as suspected, Korean security researchers have been somewhat isolated from the mainstream.

Our approach has a lot of potential. First of all, it can show the dynamics of publication trend in a given research group by comparing their scientific productions periodically. Also, we can explain the similarity/gap between the intended research group's the scientific contributions and the world leaders' those in a field.

\begin{acknowledgement}
The authors would like to thank Ross Anderson for his careful attention and insightful comments.
\end{acknowledgement}

\bibliographystyle{plain}
\bibliography{ComparativeAnalysis}

\section*{Appendix}
\appendix

\section{Centrality metrics}
\label{appendix: Centrality metrics}

\subsection{\emph{Degree}}
In a ${\bf J}$-projected graph, the \emph{degree} of a node approximately measures how many authors frequently publishes articles in the node (journal or proceeding) since the adjacent edge of the node means that an author published at least an article in this node. 

\subsection{\emph{Closeness}}
However, \emph{degree} has a shortcoming since it only takes into account the immediate edges that a node has, rather than edges to all others. Moreover, \emph{degree} do not capture the characteristics of weighted graphs. Therefore we additionally consider \emph{closeness} which focuses on the geodesic distance of a node to all others in the network. The closeness of a node $v$, $c(v)$, is computed as follows~\cite{Newman03:ComplexNetwork}:
\[
c(v) = \frac{1}{\sum_{u \in {V}}distance(v, u)}
\]
\emph{Closeness} centrality focuses on the extensibility of influence over the entire network. In a ${\bf J}$-projected graph, \emph{Closeness} measures how close all other journals/proceedings in the network are located from a given journal (or proceeding).

\subsection{\emph{Betweenness}}
The other important centrality measure is \emph{betweenness}. Let $\sigma_{st}$ denote the number of the shortest paths from $s \in {V}$ to $t \in {V}$ where $\sigma_{ss} = 1$. Let $\sigma_{st}(v)$ denote the number of shortest paths from $s \in {V}$ to $t \in V$ passing through $v \in V$. The betweenness of a node $v$, $b(v)$, is computed as follows \cite{Freeman77:SocialNetwork, Brandes01:ComplexNetwork}:
\[
b(v) = \sum_{s \neq v \in {\bf V}} \sum_{t \neq v \in {\bf V}} \frac{\sigma_{st}(v)}{\sigma_{st}}.
\]
\emph{Betweenness} is a measure of the extent to which a node lies on the paths between others. This measure favours nodes that join communities (dense sub-networks), rather than nodes that lie inside a community. In a ${\bf J}$-projected graph, journals/proceedings with high \emph{betweeness} are connectors between separate journals/proceedings groups (depending on levels or topics).

\clearpage
\section{Central journals/proceedings in the example}
\label{appendix: Central journals/proceedings in the example}

\begin{figure*}[!h]
\centering
\begin{tabular}{c c}
\includegraphics[scale=0.25]{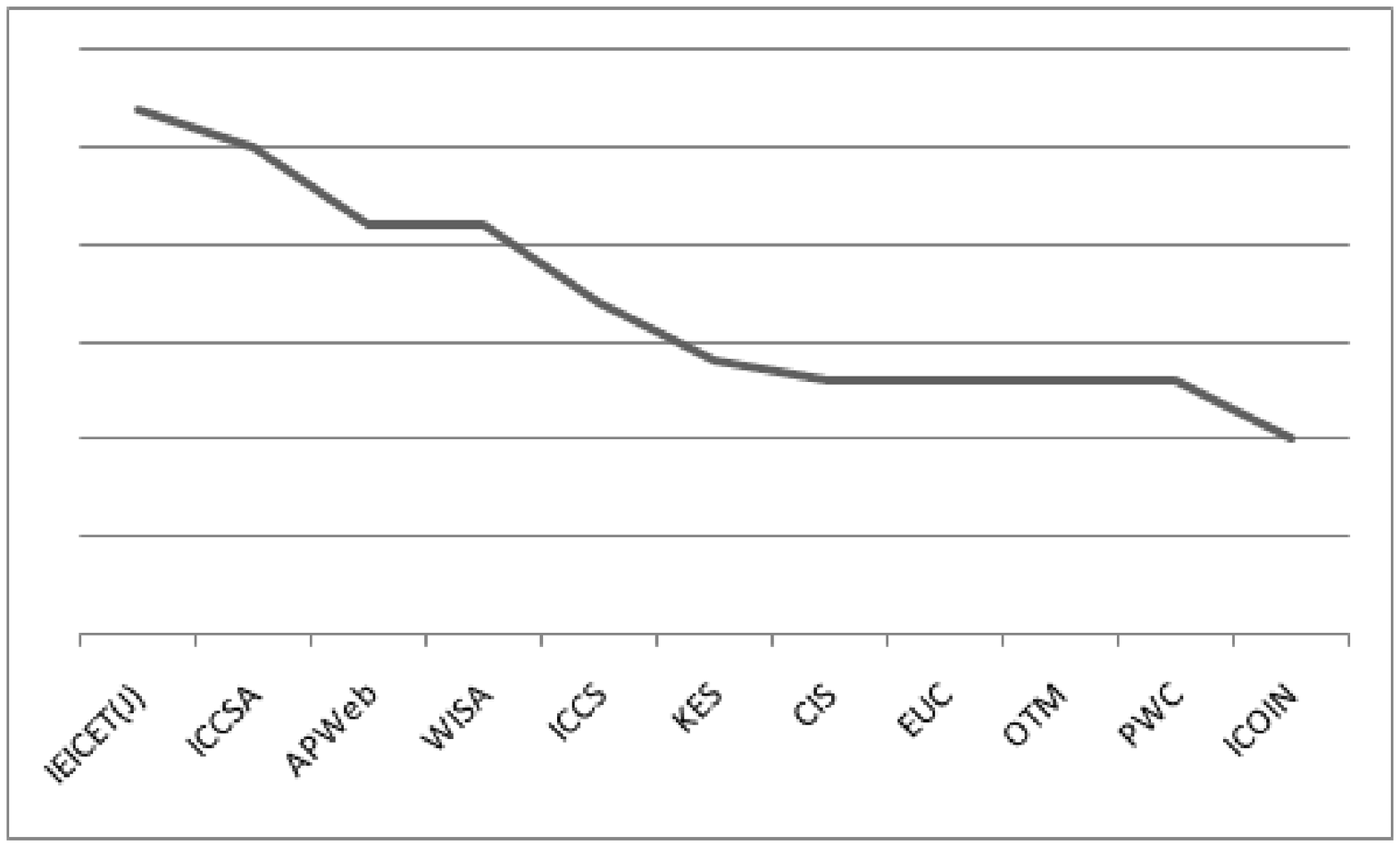}&
\includegraphics[scale=0.25]{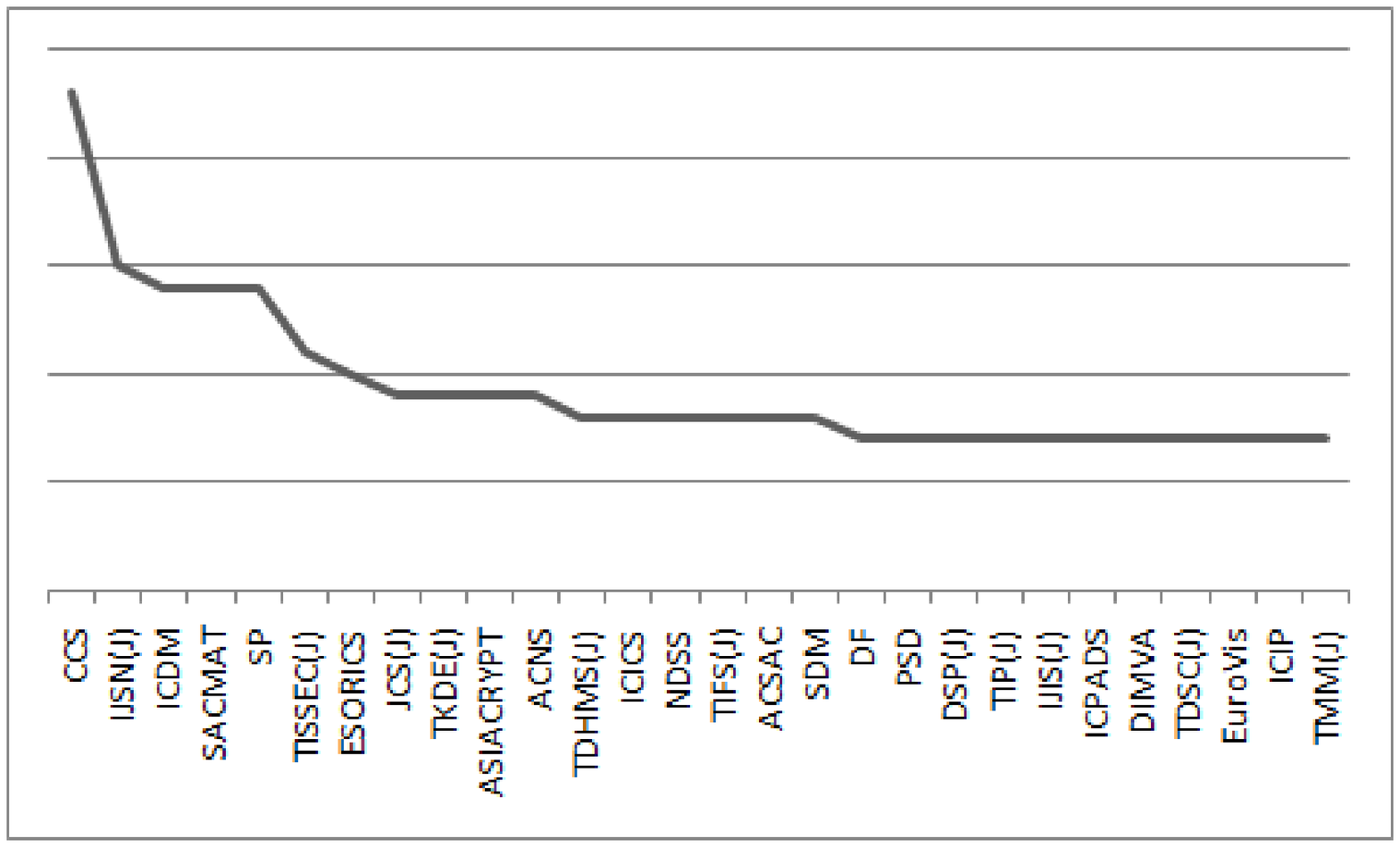}\cr
(a) Korean researchers (2006)&
(b) Global researchers (2006)\cr
\includegraphics[scale=0.25]{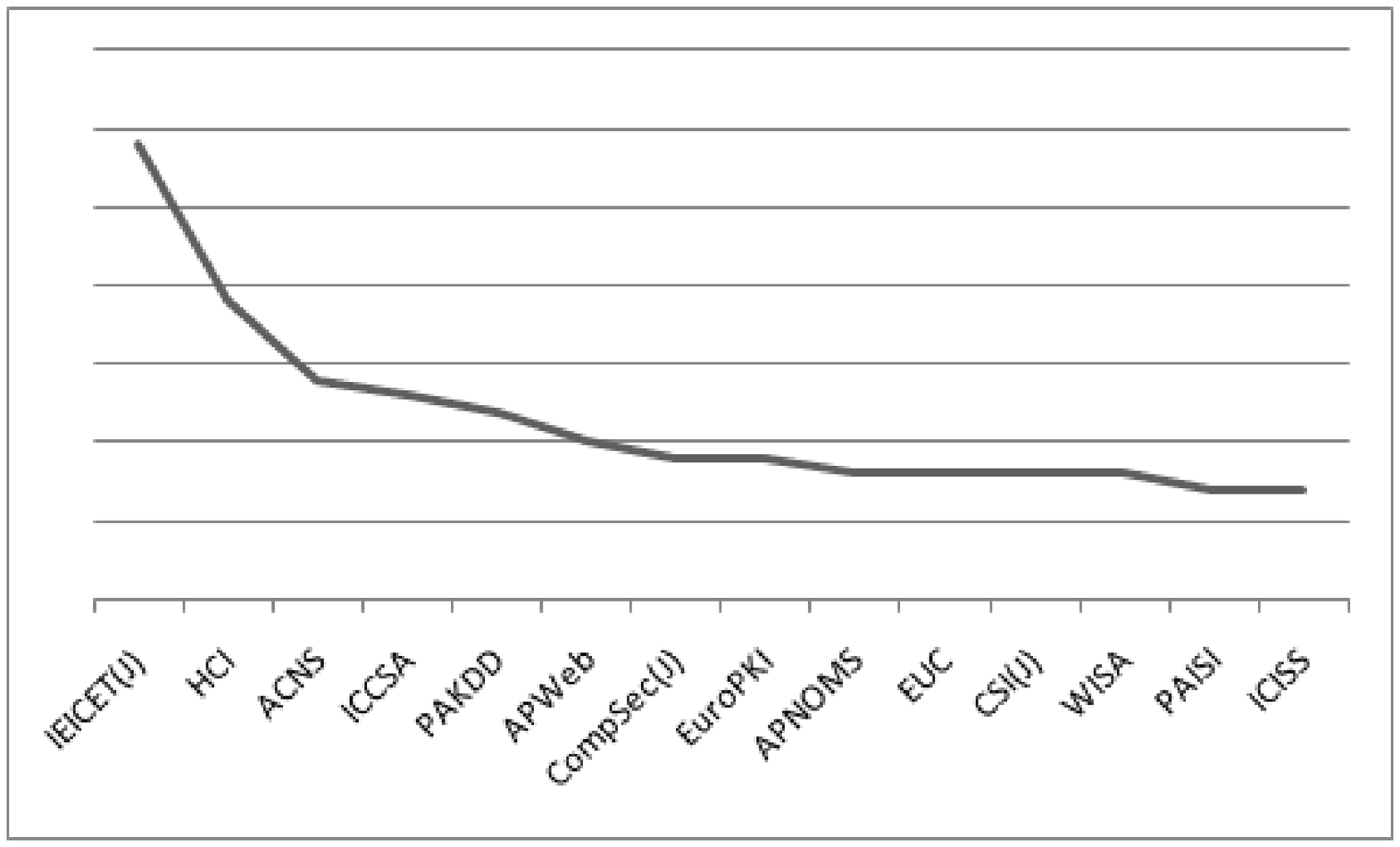}&
\includegraphics[scale=0.25]{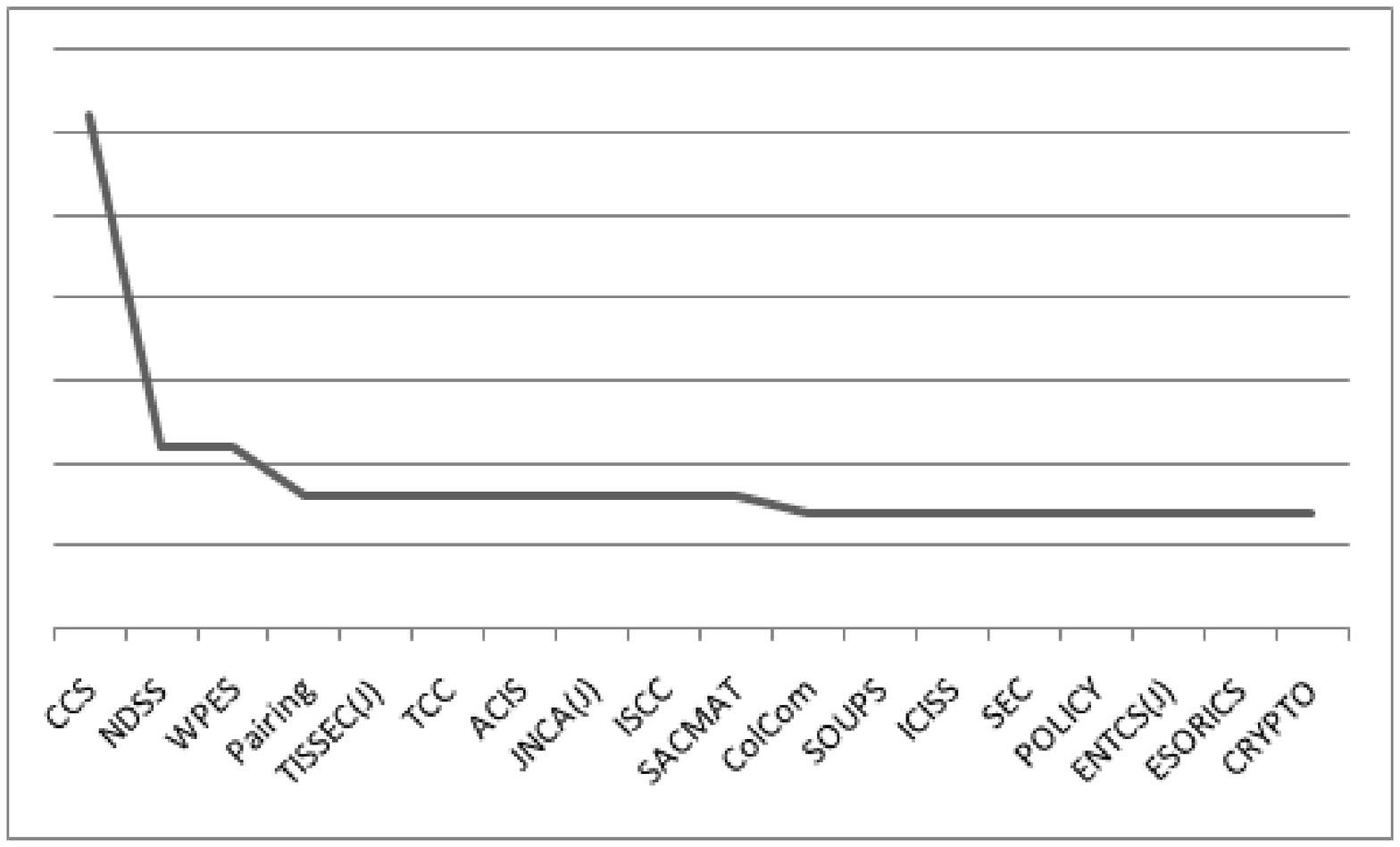}\cr
(c) Korean researchers (2007)&
(d) Global researchers (2007)\cr
\includegraphics[scale=0.25]{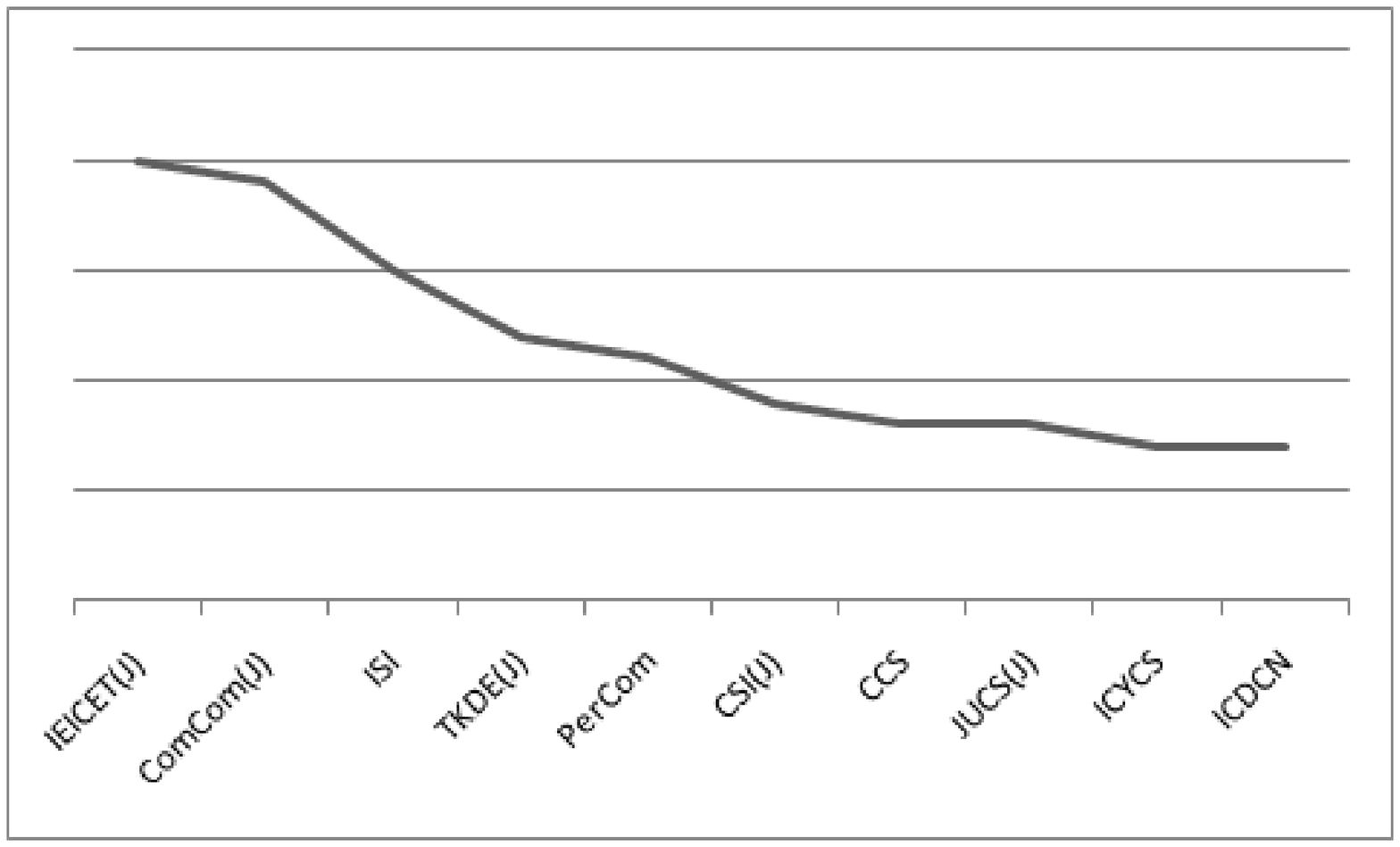}&
\includegraphics[scale=0.25]{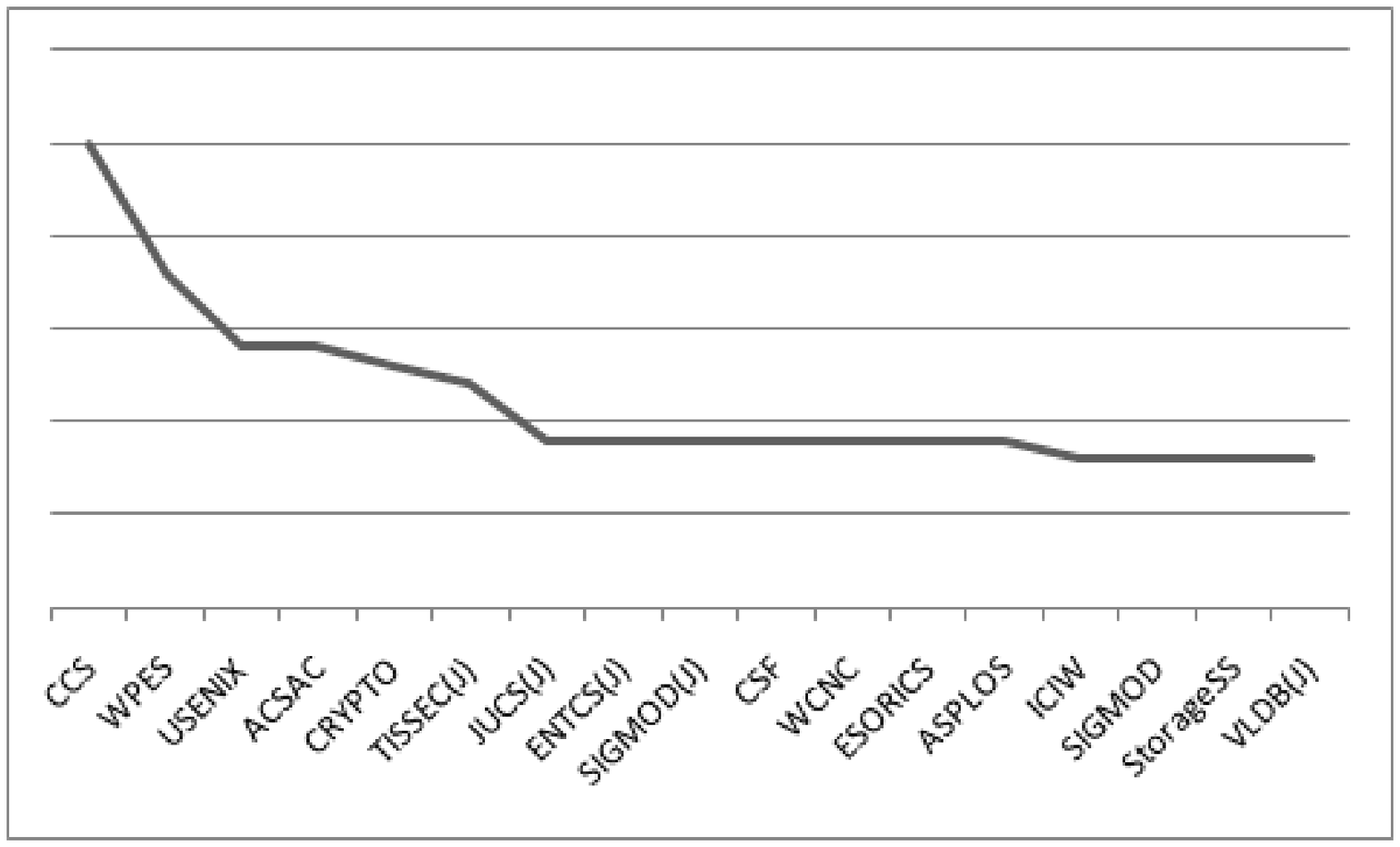}\cr
(e) Korean researchers (2008)&
(f) Global researchers (2008)\cr
\end{tabular}
\caption{\emph{degree}-central nodes in the projected graphs}
\label{fig: High degree nodes in the projected graphs}
\end{figure*}

\begin{figure*}
\centering
\begin{tabular}{c c}
\includegraphics[scale=0.25]{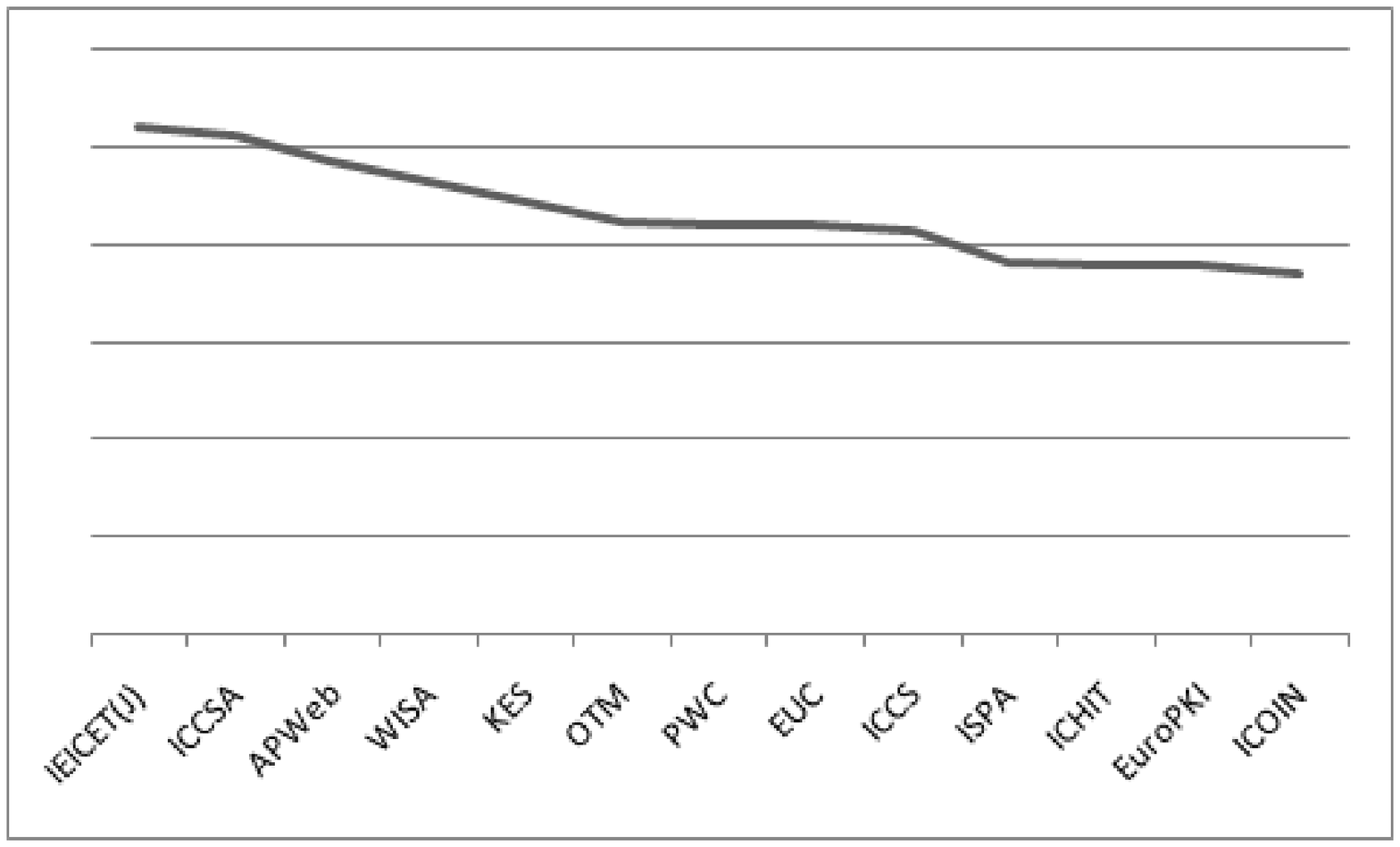}&
\includegraphics[scale=0.25]{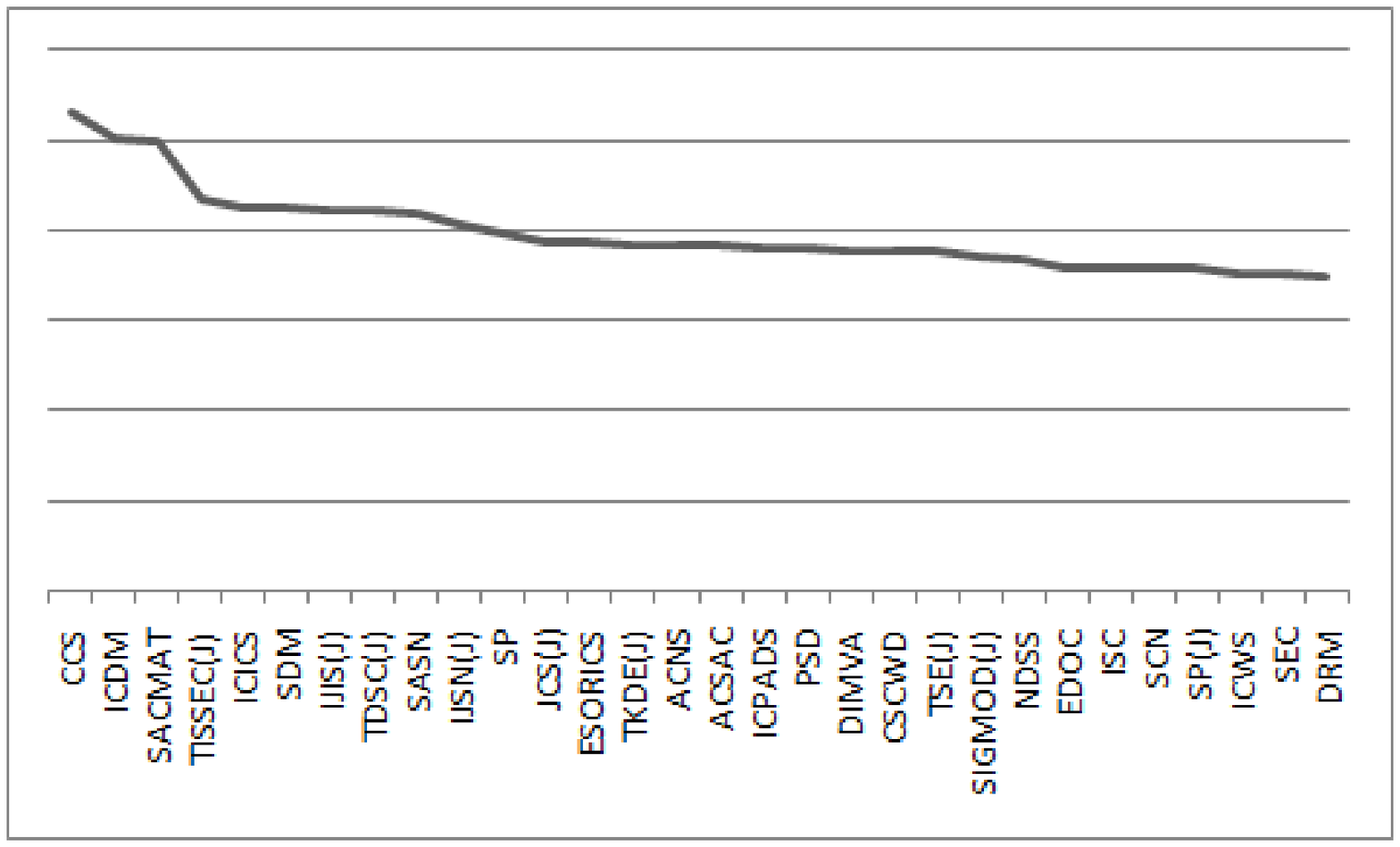}\cr
(a) Korean researchers (2006)&
(b) Global researchers (2006)\cr
\includegraphics[scale=0.25]{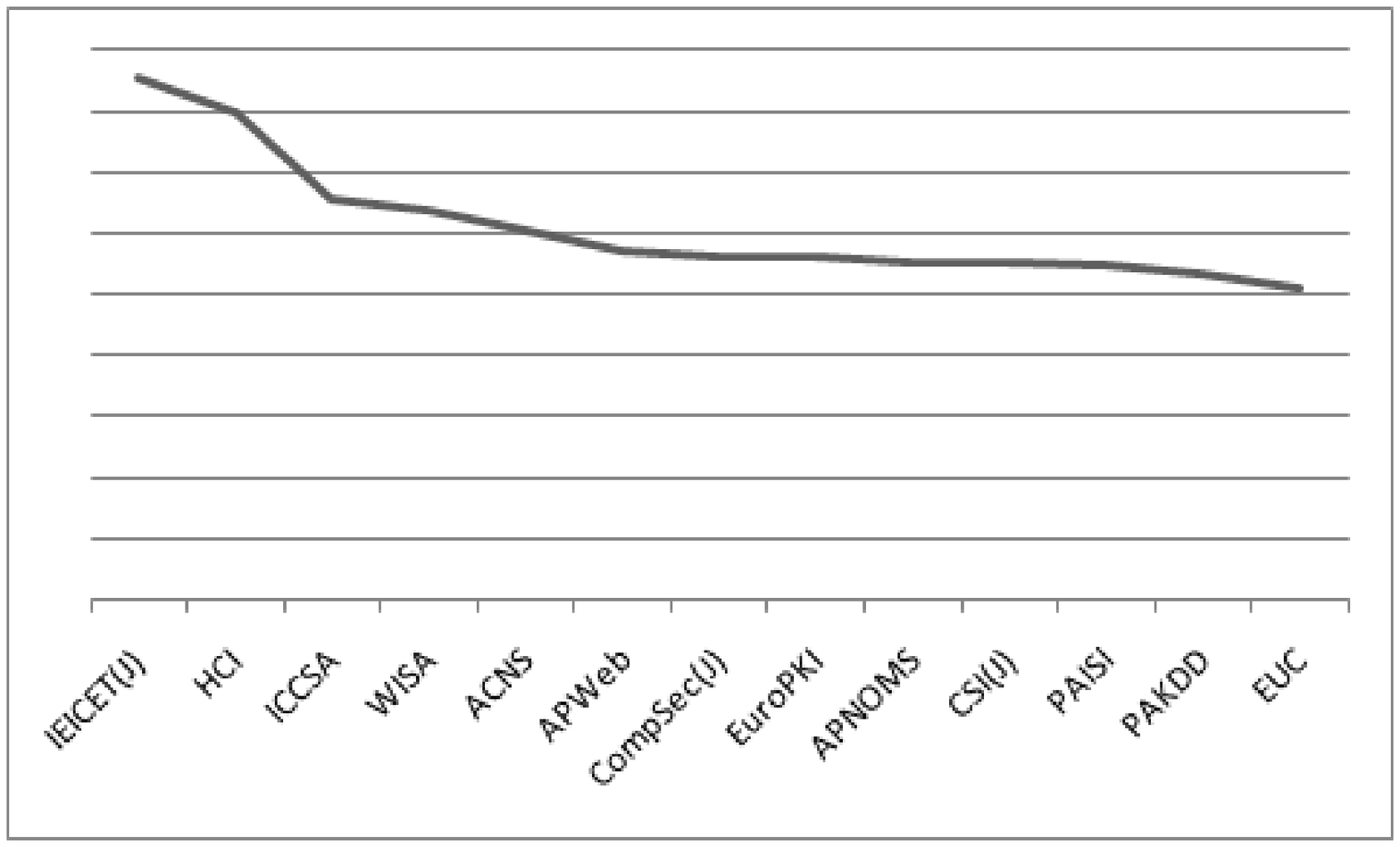}&
\includegraphics[scale=0.25]{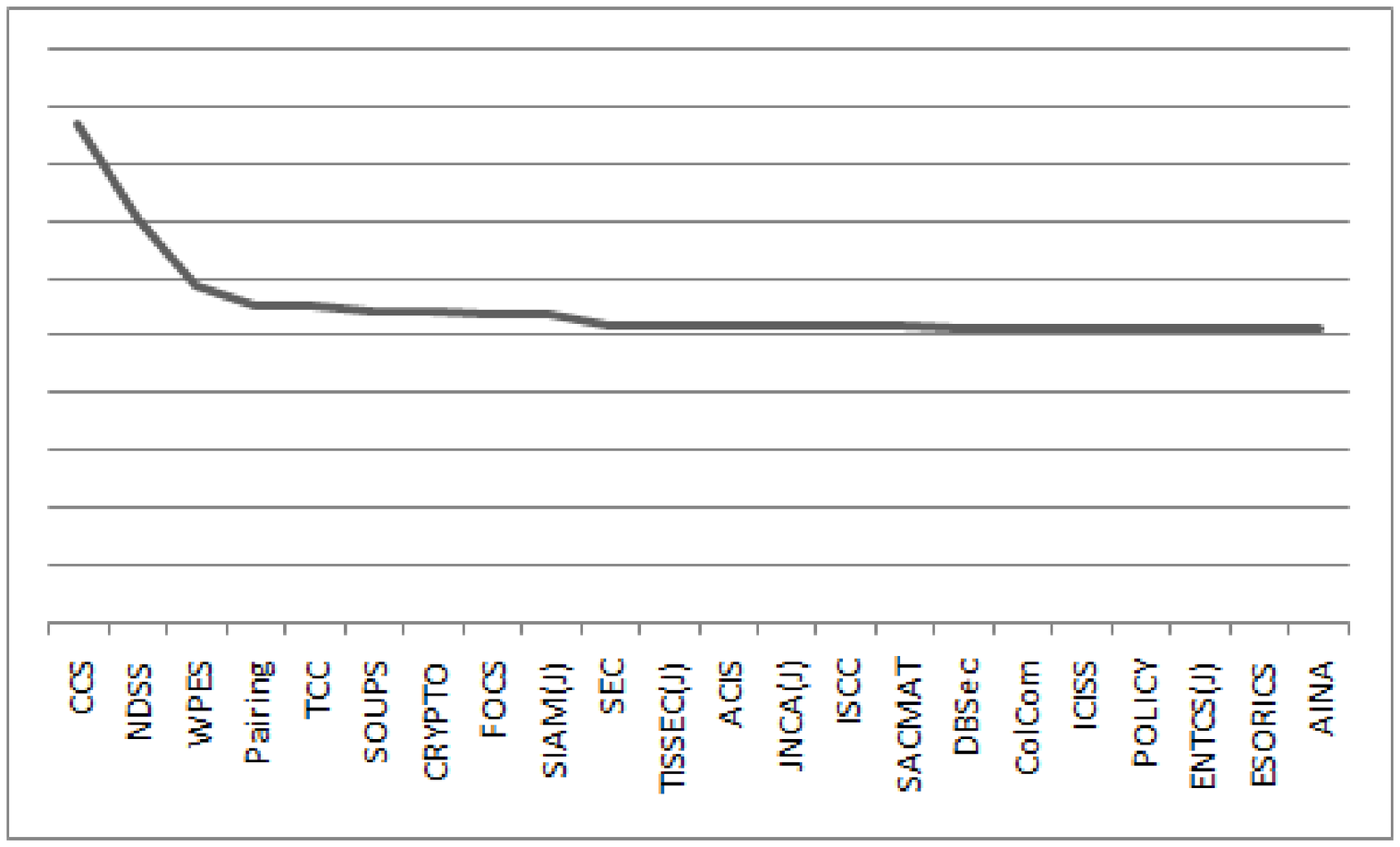}\cr
(c) Korean researchers (2007)&
(d) Global researchers (2007)\cr
\includegraphics[scale=0.25]{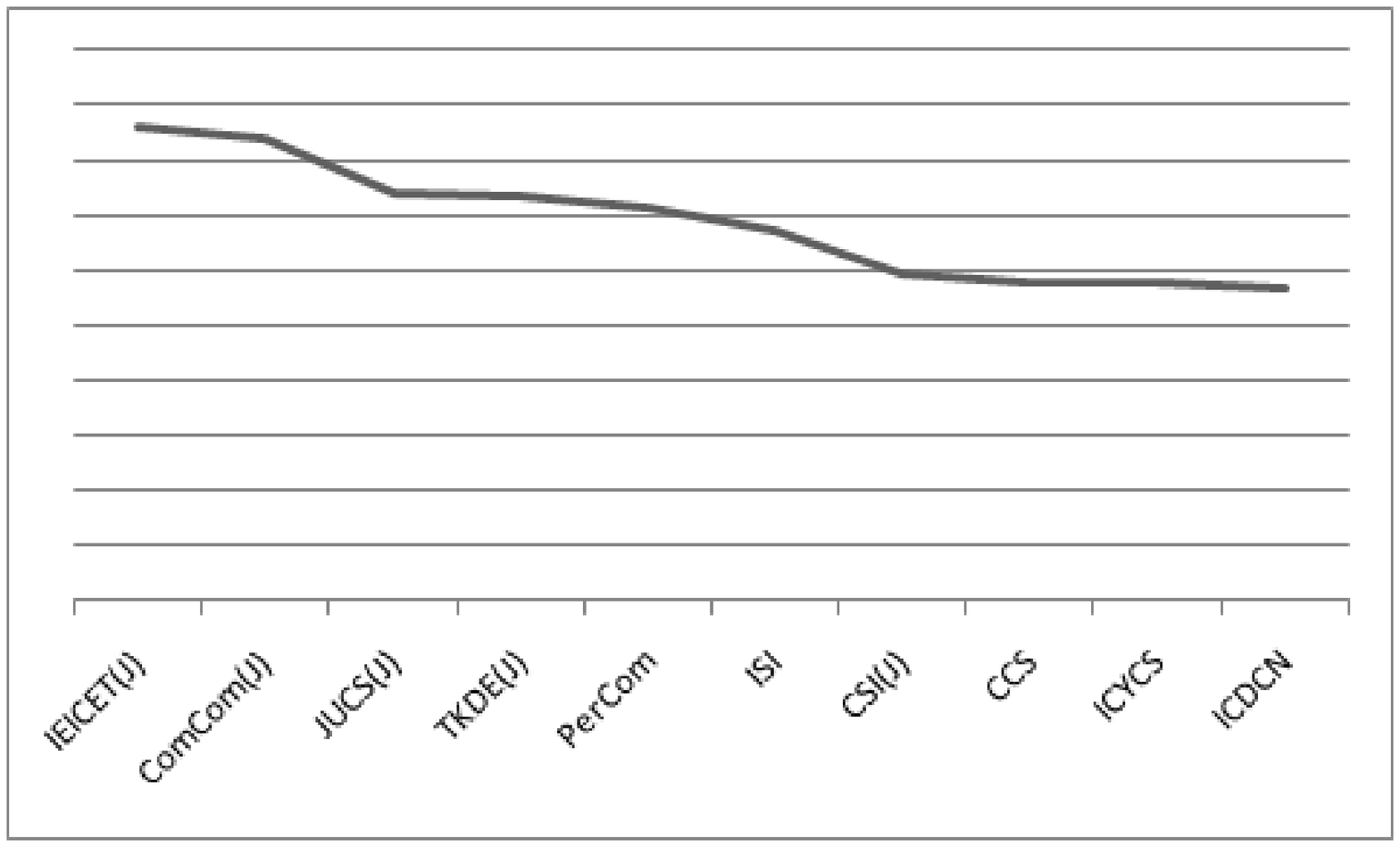}&
\includegraphics[scale=0.25]{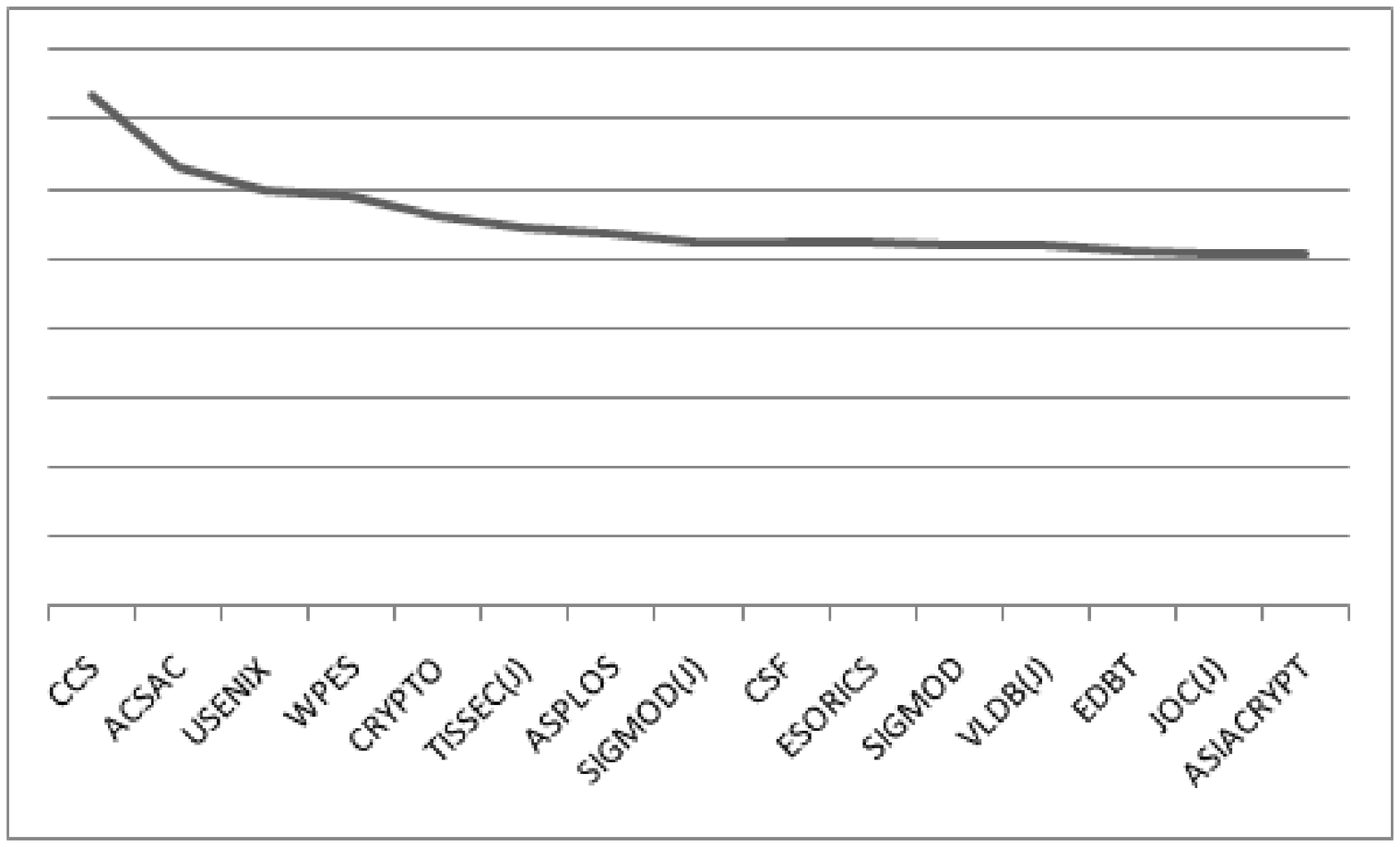}\cr
(e) Korean researchers (2008)&
(f) Global researchers (2008)\cr
\end{tabular}
\caption{\emph{closeness}-central nodes in the projected graphs}
\label{fig: High closeness nodes in the projected graphs}
\end{figure*}

\begin{figure*}
\centering
\begin{tabular}{c c}
\includegraphics[scale=0.25]{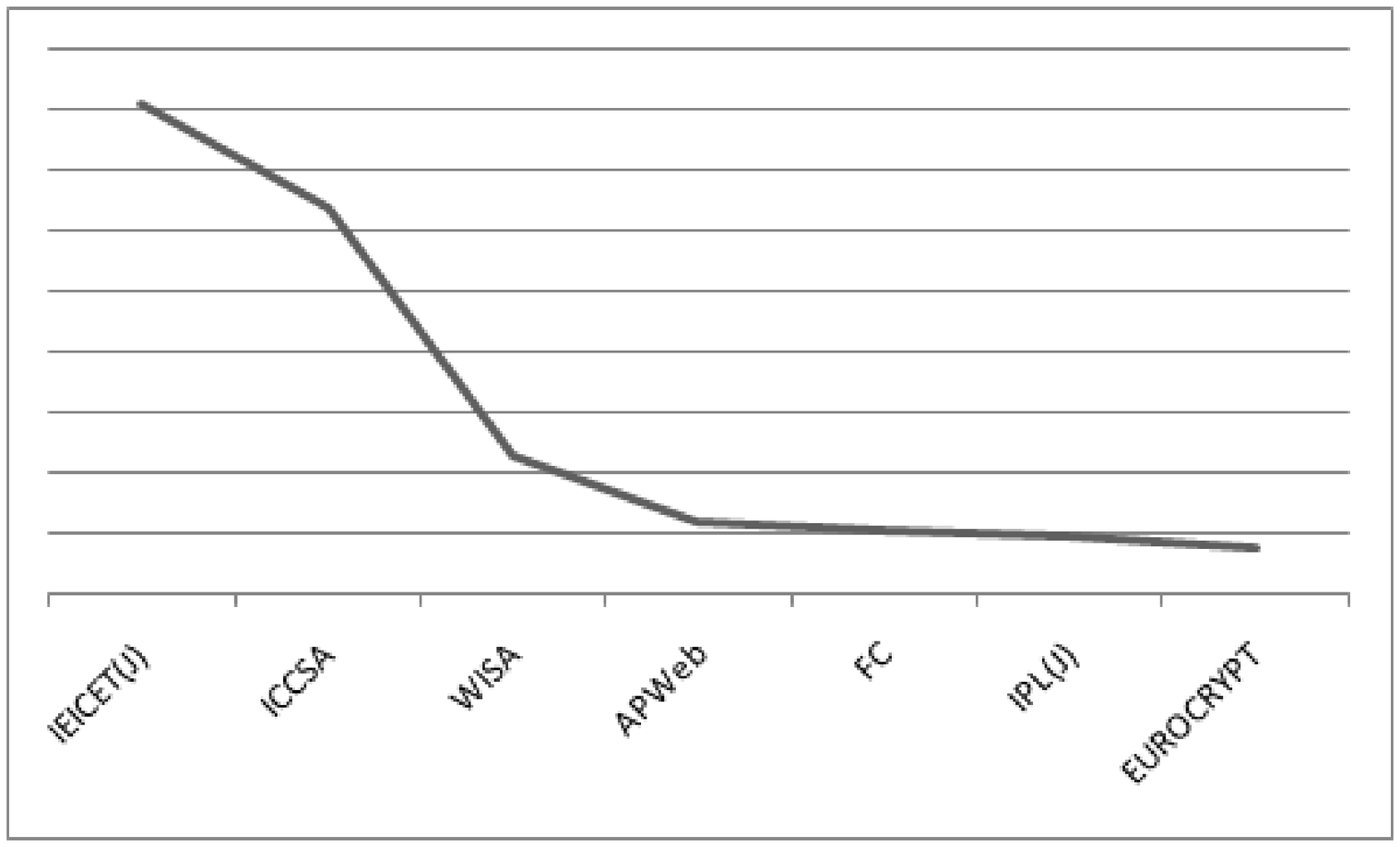}&
\includegraphics[scale=0.25]{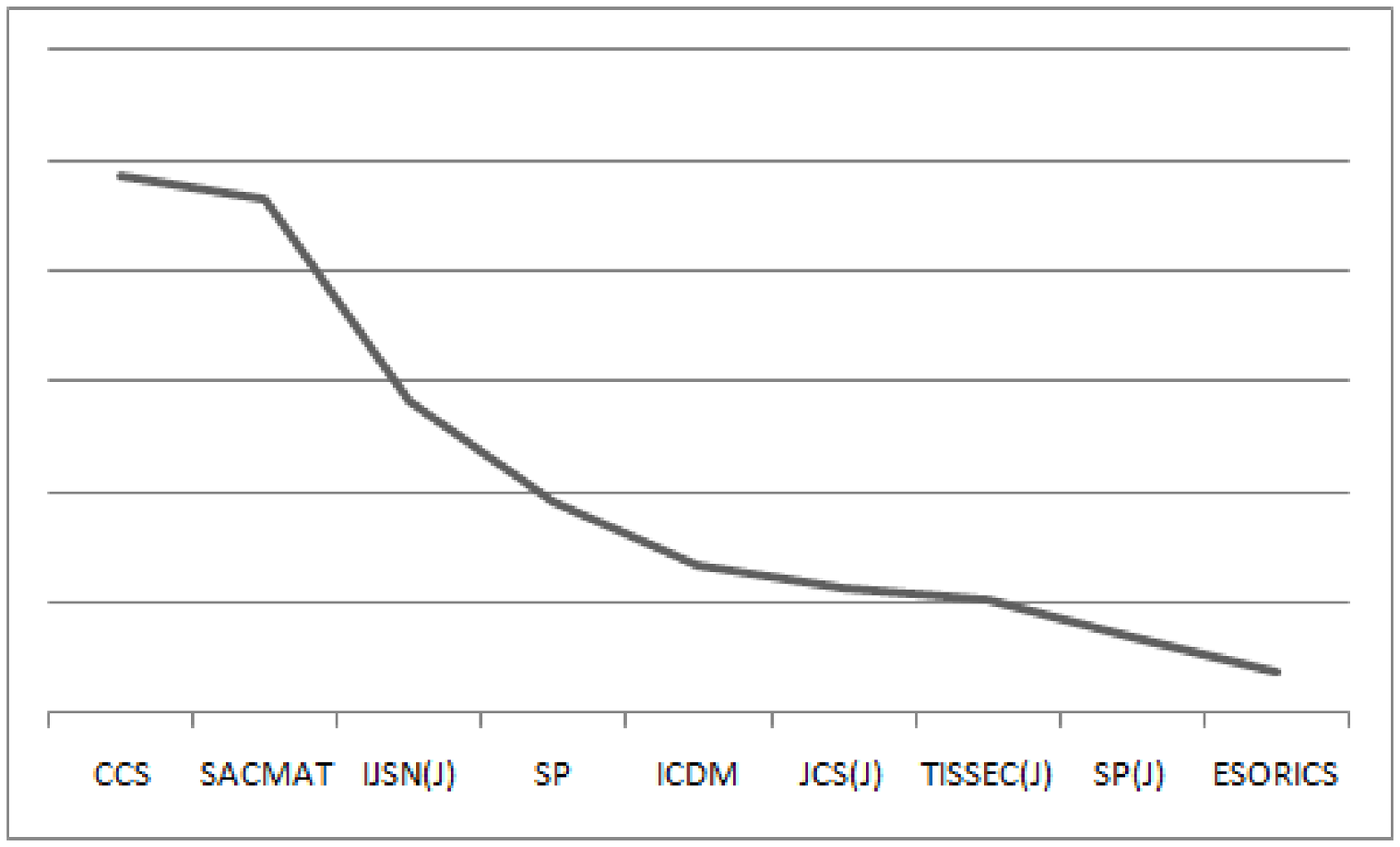}\cr
(a) Korean researchers (2006)&
(b) Global researchers (2006)\cr
\includegraphics[scale=0.25]{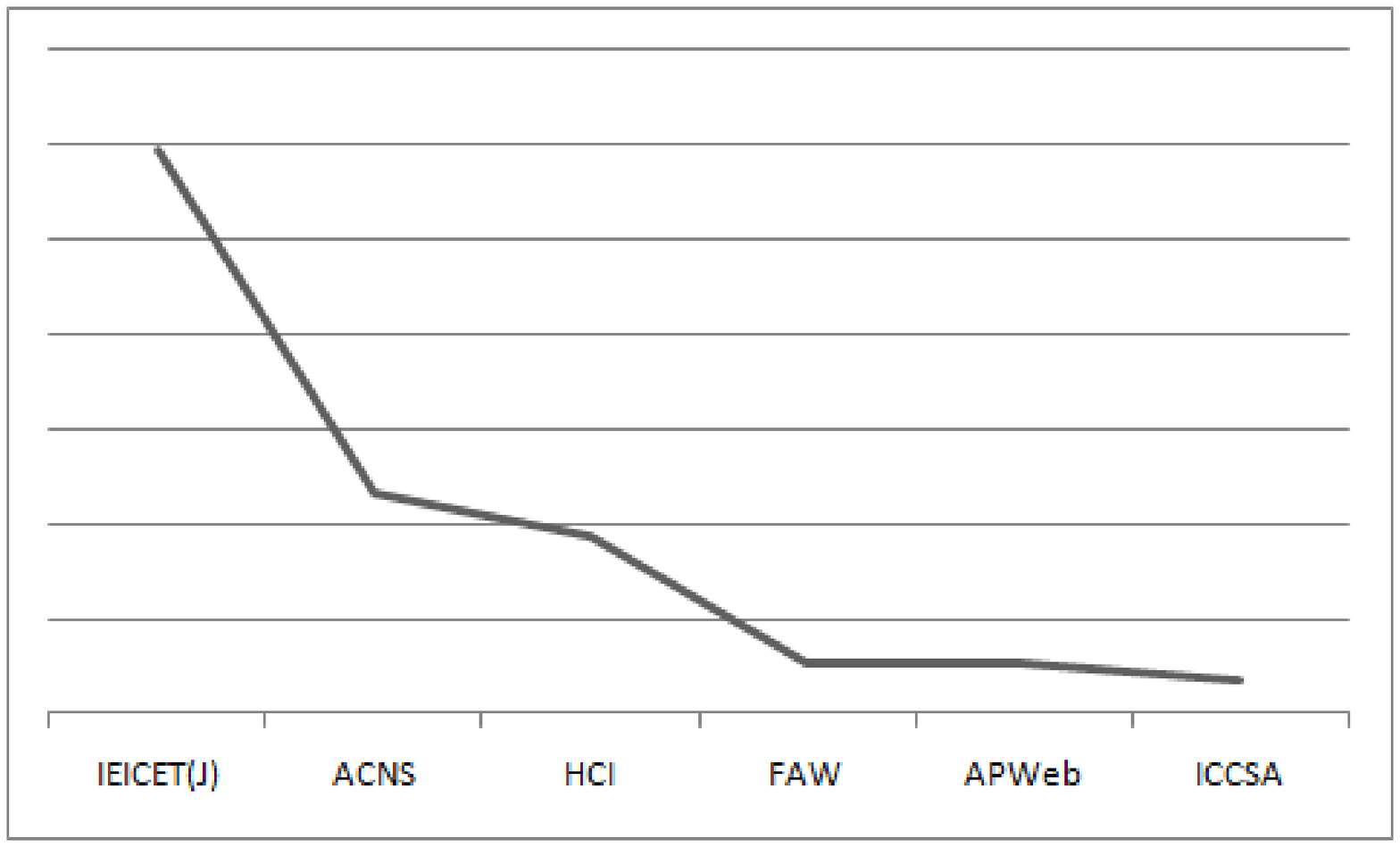}&
\includegraphics[scale=0.25]{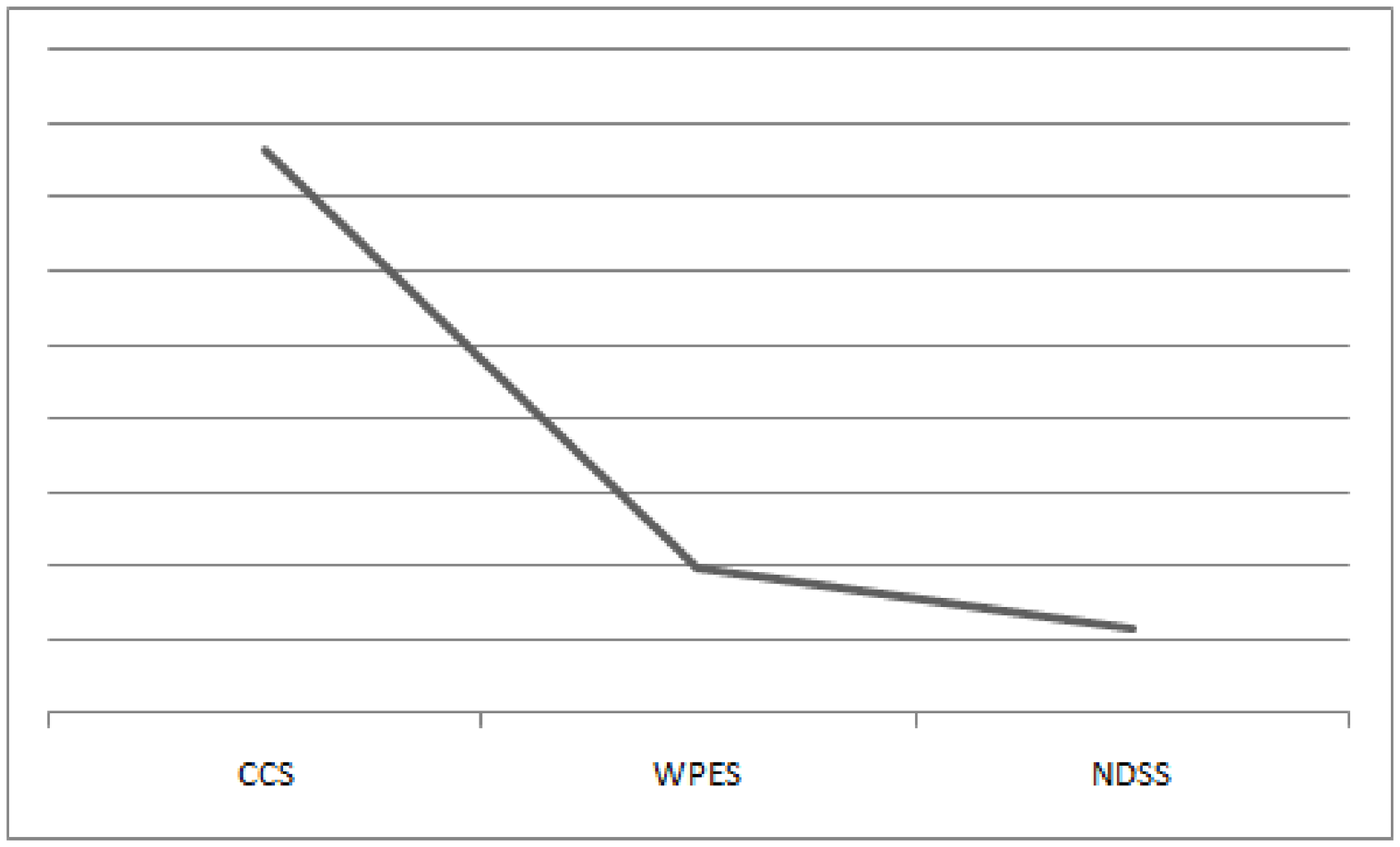}\cr
(c) Korean researchers (2007)&
(d) Global researchers (2007)\cr
\includegraphics[scale=0.25]{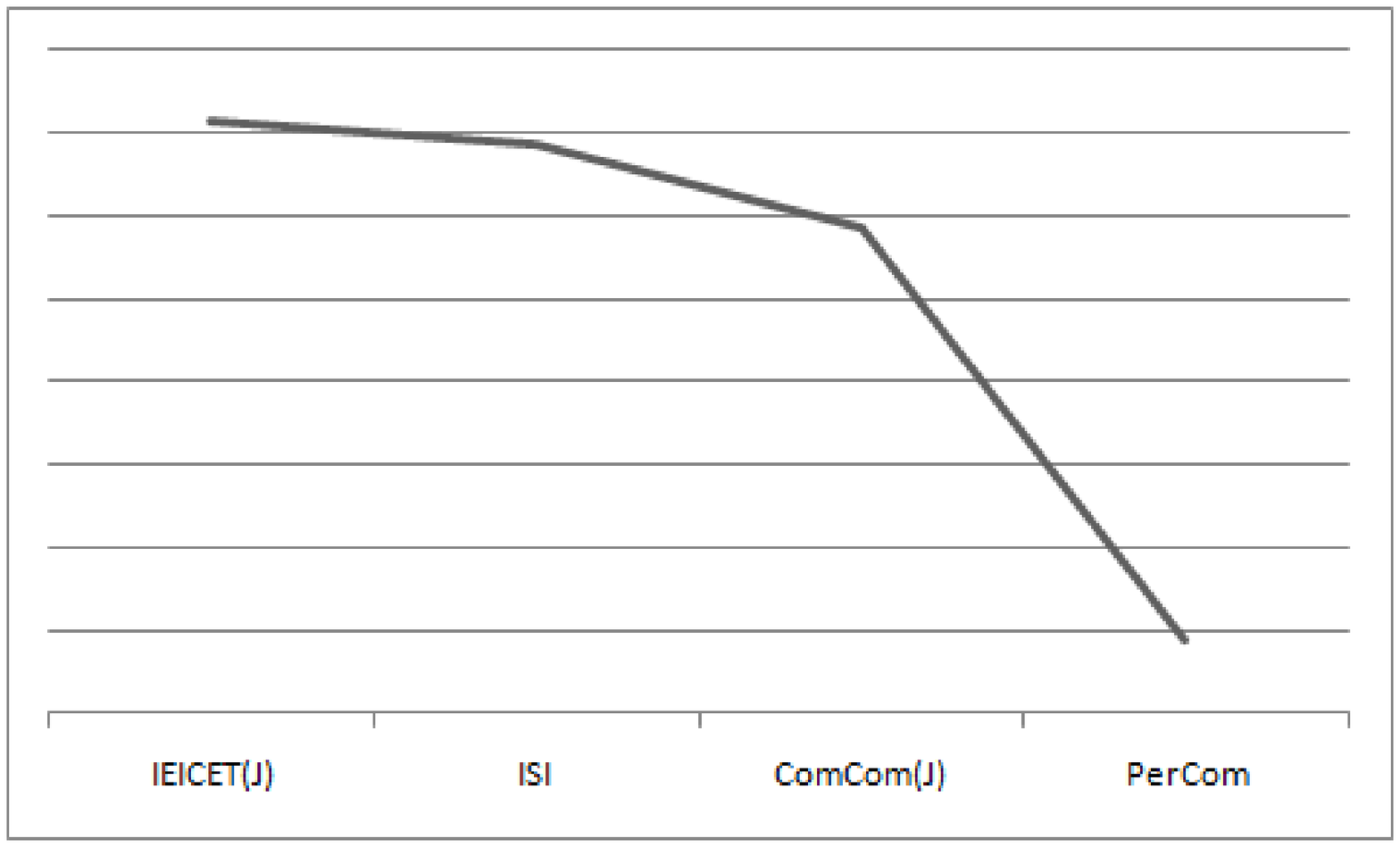}&
\includegraphics[scale=0.25]{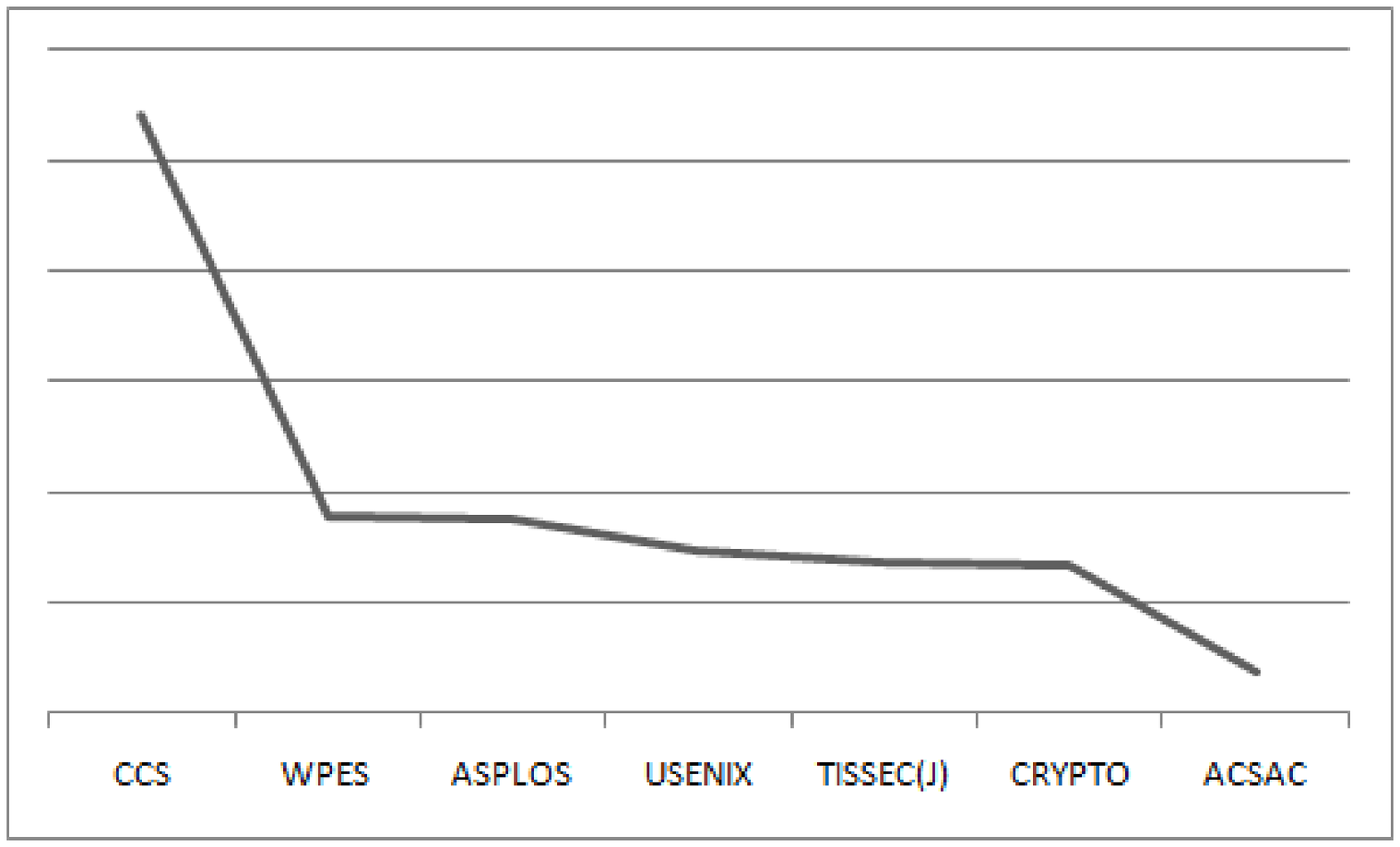}\cr
(e) Korean researchers (2008)&
(f) Global researchers (2008)\cr
\end{tabular}
\caption{\emph{betweenness}-central nodes in the projected graphs}
\label{fig: High betweenness nodes in the projected graphs}
\end{figure*}

\end{document}